\documentclass[11pt,twocolumn,aps,prb,reprint,superscriptaddress,longbibliography] {revtex4-1}
\usepackage{pst-node,graphicx,braket,amsmath,amssymb,ulem}
\usepackage[colorlinks=true,linkcolor=blue]{hyperref}

\begin{document}

\author{Dmitry A. Svintsov}
\affiliation{Center for Photonics and 2D Materials, Moscow Institute of Physics and Technology, Dolgoprudny 141700, Russia}
\email{svintcov.da@mipt.ru}
\author{Georgy V. Alymov}
\affiliation{Center for Photonics and 2D Materials, Moscow Institute of Physics and Technology, Dolgoprudny 141700, Russia}

\title{Refraction laws for two-dimensional plasmons}

\begin{abstract}
Despite numerous applications of two-dimensional plasmons for electromagnetic energy manipulation at the nanoscale, their quantitative refraction and reflection laws (analogs of Fresnel formulas in optics) have not yet been established. This fact can be traced down to the strong non-locality of equations governing the 2d plasmon propagation. Here, we tackle this difficulty by direct solution of plasmon scattering problem with Wiener-Hopf technique. We obtain the reflection and transmission coefficients for 2d plasmons at the discontinuity of 2d conductivity at arbitrary incidence angle, for both gated and non-gated 2d systems. At a certain incidence angle, the absolute reflectivity has a pronounced dip reaching zero for gated plasmons. The dip is associated with wave passage causing no dynamic charge accumulation at the boundary. For all incidence angles, the reflection has a non-trivial phase different from zero and $\pi$.

\end{abstract}

\maketitle

Quantitative laws of wave reflection from a boundary between dissimilar media play a fundamental role in physics. In electrodynamics, such relations are known as Fresnel’s formulas~\cite{fresnel_augustin_jean_2021_4561712} and represent an indispensible tool for design of any optical element, be it a cavity, a polarizer, or anti-reflection coating. Similar laws can be found in acoustics of gases, liquids and solids~\cite{landau2013fluid}. In quantum mechanics, the problem of reflection and transmission at a potential step is a primary tool to demonstrate the wave-like nature of elementary particles.

Electromagnetically thin conductive media, be it 2d materials, quantum wells, or inversion layers in semiconductors, support a special type of electromagnetic waves known as two-dimensional plasmons~\cite{Stern1967,Allen1977,Chaplik1972}. At realistic densities of charge carriers, they can be confined by $\sim10^2$ times compared to free-space electromagnetic wavelength in vacuum~\cite{Iranzo2018}. This fact motivates their application for compact light detectors~\cite{Muravev2016a,Knap2002,Bandurin2018d} and sources~\cite{ElFatimy2010,Tombet_amplification}, as well as for observation of zero-point electromagnetic fluctuation phenomena at the macro-scale~\cite{Orgiu2015}. 

Given the above motivation, it is surprising that quantitative Fresnel-type laws of 2d plasmon reflection haven’t been yet derived. The complexity of such derivation stems from strong non-locality of dynamic equations governing wave propagation. As a result of non-locality, wave-like solutions break down at the interface of two conductive media. A conventional scheme of reflectance and transmittance derivation based on matching conditions fails. A number of works dealt with approximate reflection laws for 2d plasmons using numerical techniques~\cite{Fogler2018,Siaber2019,Semenenko_scattering,Peres_PRB_Scattering} and simulators~\cite{Farajollahi2016}, yet exact expressions for reflectance and transmittance have not been obtained.

 Here, we resolve this complexity by a direct solution of scattering integral equation in piecewise-uniform 2d medium with the Wiener-Hopf technique. It is widely applied to diffraction problems at semi-infinite interfaces (wedges~\cite{Senior_WienerHopf}, waveguide terminations~\cite{Waveguide_diffraction} and others). It was used many years ago in the problem of surface wave reflection at the normal incidence between metals with dissimilar surface impedances~\cite{Kay1959} and, quite recently, to 2d plasmon incident normally to the boundary between two regions of graphene with different conductivity~\cite{Rejaei2015}. A problem of inclined incidence is more complicated due the presence of two non-trivial field components, wherein two coupled integral equations are formed~\cite{daniele2014wiener}. The Wiener-Hopf method is generally inapplicable in these situations~\cite{zabolotnykh2016edge}. The latter complexity is resolved in the quasi-static approximation. Such approximation is successfully used to describe the spectrum of edge magnetoplasmons~\cite{volkov1988edge} and similar waves~\cite{2Component_emps,Petrov_IEBP,Lozovik-Sokolik}. 

We obtain a full analytical solution for reflection and transmission of 2d plasmon at the interface between 2d systems with different conductivities at arbitrary incidence angle $\alpha $. In addition to universal total internal reflection, we find a certain angle $\alpha^*$ at which the reflection is minimized. The reflection falls completely to zero if the wave propagates in the presence of ground plane (gated plasmon). This phenomenon may look similar to Brewster effect in optics, but has a different origin. At this angle $\alpha^*$, the incident and transmitted waves cause no accumulation of charge at the interface, hence, no physical reason for reflection appears. In the case of non-gated plasmon, the reflection coefficient has a non-trivial phase shift which becomes large in the case of gliding incidence.

\begin{figure}
    \includegraphics[width = 0.95\linewidth]{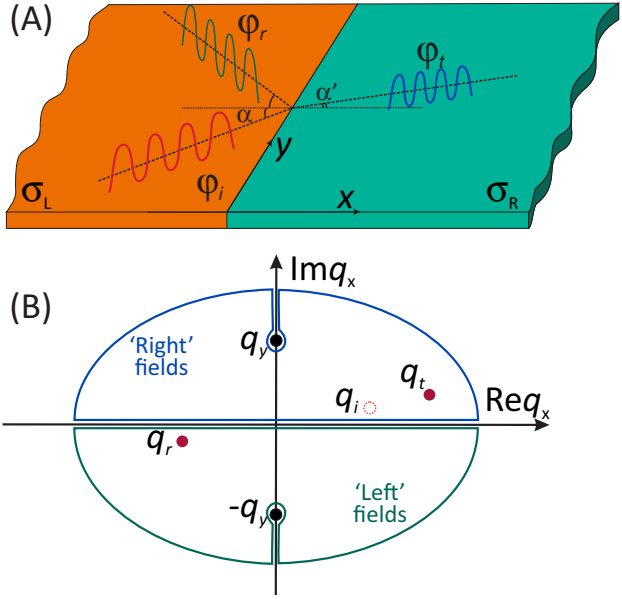}
    \caption{\label{fig1} (A) Schematic of the problem: a 2d plasma wave (red) is incident from the left on the conductivity step from $\sigma_L$ to $\sigma_R$, causing a reflected wave (green) and a transmitted wave (blue). Incidence angle is $\alpha$, refraction angle is $\alpha'$ (B) Analytic structure of the Fourier-transformed scattering equation. The 2d dielectric functions  $\varepsilon_{L/R}(q_x)$ have branch cuts starting at $\pm iq_y$ and running to $\pm i\infty$. They have simple zeros at wave vectors of the incident, transmitted and reflected waves, $q_x = \{q_i,\, q_t,\, q_r\}$. The 'incident' zero is compensated by the pole of Fourier-transformed incident potential (shown with hollow circle).}
\end{figure}

We proceed to the solution of the scattering problem for 2d plasmons schematically shown in Fig.~\ref{fig1} a. A plasma wave is incident from the left 2d section with conductivity $\sigma_L$ at the boundary with right 2d section with conductivity $\sigma_R$ at angle $\alpha$, causing a reflected (r) and transmitted (t) waves. All wave characteristics (potential $\varphi $, current density $\mathrm{j}$) are harmonically varying in time as ${{e}^{-i\omega t}}$, this time-dependent term will be skipped. The frequency dependence of conductivity $\sigma(\omega)$ can be arbitrary and not limited to Drude model. The only requirement is that $\sigma$ has a large positive imaginary part, such that transverse-magnetic 2d plasmons are well-defined.

The governing equation for electric potential $\Phi \left( \mathrm{r} \right)$ in the 2d plane can be presented symbolically as:
\begin{equation}
    \Phi \left( \mathrm{r} \right)=\mathcal{L}\left[ \Phi  \right],
\end{equation}
where $\mathcal{L}\left[ ... \right]$ is the integro-differential linear operator linking the potential created by charges in 2DES to the non-uniform field producing these charges:
\begin{equation}
   \mathcal{L}\left[ f \right]=\frac{1}{i\omega }\int{{{d}^{2}}\mathrm{{r}'}G\left( \mathrm{r}-\mathrm{{r}'} \right){{\nabla }_{{\mathrm{{r}'}}}}\left[ \sigma \left( {\mathrm{{r}'}} \right){{\nabla }_{{\mathrm{{r}'}}}}f\left( {\mathrm{{r}'}} \right) \right]} 
\end{equation}
Above, $G\left( \mathrm{r} \right)={{\left| \mathrm{r} \right|}^{-1}}-{{\left| {{\mathrm{r}}^{2}}+4{{d}^{2}} \right|}^{-1/2}}$ is the Green’s function of the electrostatic problem, $d$ is the distance to the screening plate (gate), $\sigma \left( \mathrm{r} \right)$ is the distribution of 2d conductivity $\sigma \left( \mathrm{r} \right)={{\sigma }_{L}}\theta \left( -x \right)+{{\sigma }_{R}}\theta \left( x \right)$.

To solve the scattering problem, we split the full potential into the incident and scattered fields $\Phi = \varphi_i + \varphi$. We choose the incident field as a semi-bounded plasma wave ${{\varphi }_{i}}\left( \mathrm{r} \right)={{\varphi }_{i}}\exp \left( i{{q}_{i}}x+i{{q}_{y}}y \right)\theta \left( -x \right)$. After such decomposition, the governing equation for scattered fields $\varphi$ takes the form:
\begin{equation}
\label{Eq-Scattering}
    \varphi \left( \mathrm{r} \right)=\left\{ \mathcal{L}\left[ {{\varphi }_{i}} \right]-{{\varphi }_{i}}\left( \mathrm{r} \right) \right\}+\mathcal{L}\left[ \varphi  \right].
\end{equation}
We recognize that the term in curly brackets is equivalent to 'external source' creating the scattered field. From now on, the solution of scattering problem for 2d plasmons will be not much different from the solution of half-plane diffraction problems under external free-space illumination that have been studied extensively~\cite{Senior_WienerHopf,Zhang_2014,Margetis_edge_diffraction,Nikulin2021}.

We apply two subsequent Fourier transforms to Eq.~\ref{Eq-Scattering}. The first one with respect to the $y$-coordinate is trivial. The emerging wave vector ${{q}_{y}}$ will be considered as an independent variable of the problem, the conserved $y$-component of plasmon momentum. Further on, we split the scattered potential into the ‘left’ and ‘right’ functions ${{\varphi }_{{{q}_{y}}}}\left( x \right)={{\varphi }_{L}}\left( x \right)\theta \left( -x \right)+{{\varphi }_{R}}\left( x \right)\theta \left( x \right)$, and apply the second Fourier transform $F\left[ \varphi  \right]\left( {{q}_{x}} \right)=\int\limits_{-\infty }^{+\infty }{\varphi \left( x \right){{e}^{-i{{q}_{x}}x}}dx}$
with respect to $x$-coordinate. This leads us to fully Fourier-transformed scattering problem:
\begin{multline}
\label{Eq-Scattering_Fourier}
  {{\varepsilon }_{L}}\left( {{q}_{x}} \right)\left[ {{\varphi }_{i}}\left( {{q}_{x}} \right)+{{\varphi }_{L}}\left( {{q}_{x}} \right) \right]+{{\varepsilon }_{R}}\left( {{q}_{x}} \right){{\varphi }_{R}}\left( {{q}_{x}} \right)=\\
  \frac{{{q}_{x}}}{\omega }G\left( q \right)\left[ {{\sigma }_{R}}-{{\sigma }_{L}} \right]\varphi_0,   
\end{multline} 
where we have introduced the effective 2d dielectric functions of the left and right media 
\begin{equation}
{{\varepsilon }_{\alpha }}\left( {{q}_{x}} \right)=1+\frac{i{{\sigma }_{\alpha }}}{\omega }{{q}^{2}}G\left( q \right), \qquad \alpha =\{L, \, R \}
\end{equation}
and the Fourier-transformed Green’s function of the electrostatic problem $G\left( q \right)=2\pi q^{-1}( 1-{{e}^{-2qd}} )$, $q=[q_{x}^{2}+q_{y}^{2}]^{1/2}$. The term on the right-hand side containing the value or real-space potential at the boundary ${\varphi_{q_y}}( 0 ) \equiv \varphi_0$ has emerged due to discontinuous electric field at the boundary.

Solution of (\ref{Eq-Scattering_Fourier}) is based on inspection of analytic properties of emerging functions in the plane of complex ${{q}_{x}}$-variable and is given in Supplementary material, section I. The main property is that two functions ${{F}_{+}}\left( {{q}_{x}} \right)$ and ${{F}_{-}}\left( {{q}_{x}} \right)$ being analytic in the upper and lower half-planes and identical in a stripe $\left| \operatorname{Im}{{q}_{x}} \right|<\delta $ should be equal to a polynomial of complex ${{q}_{x}}$. This polynomial degenerates to zero if we require finiteness of potentials at infinity. Such splitting of Eq. (\ref{Eq-Scattering_Fourier}) is quite straightforward if we know the decomposition of dielectric function ${{\varepsilon }_{\alpha }}\left( {{q}_{x}} \right)=\varepsilon _{\alpha }^{+}\left( {{q}_{x}} \right)\varepsilon_{\alpha }^{-}\left( {{q}_{x}} \right)$ into the functions analytic and free of zeros in the upper (+) and lower (-) half-planes. In any case, it can be achieved with general formula:
\begin{equation}
\varepsilon_{\alpha}^{\pm}( q_x ) = \exp\left\{\pm\frac{1}{2\pi i} \int_{-\infty}^{+\infty}{\frac{\ln \varepsilon_\alpha (u)du}{u-q_x \pm i 0^+ }}\right\},   
\end{equation}
while alternative approaches and semi-analytical formulas can also be available (see Supplementary Material, section II).

The result of splitting for scattering equation (\ref{Eq-Scattering_Fourier}) reads as
\begin{multline}
 \left[ {{M}_{+}}\left( {{q}_{x}} \right)-{{M}_{+}}\left( {{q}_{i}} \right) \right]{{\varphi }_{i}}\left( {{q}_{x}} \right)+{{M}_{+}}\left( {{q}_{x}} \right){{\varphi }_{L}}\left( {{q}_{x}} \right)-i\frac{\varphi_0}{2}{{L}_{+}}\left( {{q}_{x}} \right)= \\ 
  -{{M}_{+}}\left( {{q}_{i}} \right){{\varphi }_{i}}\left( {{q}_{x}} \right)-{{M}_{-}}\left( {{q}_{x}} \right){{\varphi }_{R}}\left( {{q}_{x}} \right)+i\frac{\varphi_0}{2}{{L}_{-}}\left( {{q}_{x}} \right),
\end{multline}
 \begin{gather}
{{M}_{+}}\left( {{q}_{x}} \right)=\frac{\varepsilon _{L}^{+}\left( {{q}_{x}} \right)}{\varepsilon _{R}^{+}\left( {{q}_{x}} \right)},{{M}_{-}}\left( {{q}_{x}} \right)=\frac{\varepsilon _{R}^{+}\left( {{q}_{x}} \right)}{\varepsilon _{-}^{+}\left( {{q}_{x}} \right)}, \\ 
 {{L}_{\pm }}\left( {{q}_{x}} \right)=\pm \frac{{{M}_{\pm }}\left( {{q}_{x}} \right)}{{{q}_{x}}\pm i{{q}_{_{y}}}}\pm \frac{{{M}_{\pm }}\left( {{q}_{x}} \right)-{{M}_{\pm }}\left( \pm i{{q}_{y}} \right)}{{{q}_{x}}\mp i{{q}_{_{y}}}}\mp \frac{{{M}_{\mp }}\left( \mp i{{q}_{y}} \right)}{{{q}_{x}}\pm i{{q}_{_{y}}}}.
\end{gather}

The left- and right-hand sides of such equation are now analytic in the upper and lower half-planes, respectively. They are identical in the stripe $\left| \operatorname{Im}{{q}_{x}} \right|<\operatorname{Im}{{q}_{i}}$, where $\operatorname{Im}{{q}_{i}}$ is the decay constant of the incident wave (can approach zero in the final result). Hence, both sides are zero identically, which yields the solution for scattering problem in the Fourier space:
\begin{multline}
\label{Eq-L}
 {{\varphi }_{L}}\left( {{q}_{x}} \right)=M^{-1}_+(q_x)\times\\
 \left\{ i\frac{\varphi_0}{2}{{L}_{+}}\left( {{q}_{x}} \right)-\left[ {{M}_{+}}\left( {{q}_{x}} \right)-{{M}_{+}}\left( {{q}_{i}} \right) \right]{{\varphi }_{i}}\left( {{q}_{x}} \right) \right\} 
 \end{multline}
\begin{multline}
\label{Eq-R}
{{\varphi }_{R}}\left( {{q}_{x}} \right)=M^{-1}_-(q_x)\times\\
\left\{ i\frac{\varphi_0}{2}{{L}_{-}}\left( {{q}_{x}} \right)-{{M}_{+}}\left( {{q}_{i}} \right){{\varphi }_{i}}\left( {{q}_{x}} \right) \right\}    .  
\end{multline}
A remaining problem is to link the real-space potential at $x=0$, $\varphi_0$, to that in the incident wave, $\varphi_i$. This is achieved by evaluating the inverse transform $\varphi_0 =\pi^{-1}\lim_{x\to 0^-} \int_{-\infty }^{+\infty }{{{\varphi }_{L}}\left( {{q}_{x}} \right) e^{-i q_x x}d{{q}_{x}}}$ and solving a simple self-consistency system. This leads to
\begin{equation}
 \varphi_0=2{{\varphi }_{i}}\frac{{{M}_{+}}\left( {{q}_{i}} \right)}{{{M}_{+}}\left( i{{q}_{y}} \right)+{{M}_{-}}\left( -i{{q}_{y}} \right)}
 \end{equation}
and completes the formal solution.

The real-space profiles of fields $\varphi_L(x)$ and $\varphi_R(x)$ are evaluated by inverse Fourier transforms of (\ref{Eq-L}) and (\ref{Eq-L}). The spatial structure of the real fields becomes transparent if we evaluate the transforms along the loops in $q_x$-plane shown in Fig.~\ref{fig1} b. The branch-cut contributions to $\varphi(x)$ would correspond to evanescent fields near the edge with non-propagating nature. The residues at the poles $q_x = q_r$ and $q_x = q_t$ would yield the amplitudes or transmitted and reflected plasmons, respectively:
\begin{widetext}
\begin{gather}
\label{Eq-reflection}
r=\frac{{{M}_{+}}\left( {{q}_{i}} \right)}{{{\left. \partial {{M}_{+}}/\partial {{q}_{x}} \right|}_{{{q}_{x}}={{q}_{r}}}}}\left\{ \frac{1}{{{q}_{r}}-{{q}_{i}}}-\frac{{{q}_{r}}}{q_{r}^{2}+q_{y}^{2}}-\frac{i{{q}_{y}}}{q_{r}^{2}+q_{y}^{2}}\frac{M_{+}^{2}\left( i{{q}_{y}} \right)-1}{M_{+}^{2}\left( i{{q}_{y}} \right)+1} \right\},\\
\label{Eq-Transmission}
t=\frac{1}{{{\left. \partial {{M}_{-}}/\partial {{q}_{x}} \right|}_{{{q}_{x}}={{q}_{t}}}}}\left\{ \frac{{{M}_{+}}\left( {{q}_{i}} \right)}{{{q}_{t}}-{{q}_{i}}}-{{M}_{+}}\left( {{q}_{t}} \right)\left[ \frac{2{{q}_{t}}}{q_{t}^{2}+q_{y}^{2}}-\frac{i{{q}_{y}}}{q_{t}^{2}+q_{y}^{2}}\frac{M_{-}^{2}\left( -i{{q}_{y}} \right)-1}{M_{-}^{2}\left( -i{{q}_{y}} \right)+1} \right] \right\}
\end{gather}
\end{widetext}
Equations \ref{Eq-reflection} and \ref{Eq-Transmission} represent the central results of this paper. To check its correctness, we note that for small separation between gate and 2DES $qd\ll 1$, the dielectric function $\varepsilon _{\alpha }^{G}\left( {{q}_{x}} \right)$ has a very simple analytic structure. Namely 
$\varepsilon _{\alpha }^{G}\left( {{q}_{x}} \right)=1-\left( q_{x}^{2}+q_{y}^{2} \right)/q_{p\alpha }^{2}$, where $q_{p\alpha }^{2}=i\omega / 4\pi d{{\sigma }_{\alpha }}$ is the absolute value of plasmon wave vector. The factorization of such dielectric function is immediately achieved:
\begin{equation}
\label{Eq-Gated-Factorization}
 \varepsilon _{\alpha }^{G}\left( {{q}_{x}} \right)=\frac{\sqrt{q_{p\alpha }^{2}-q_y^2}-{q_x}}{{q_{p\alpha }}}\frac{\sqrt{q_{p\alpha }^2-q_y^2}+{q_x}}{{q_{p\alpha }}}
 \end{equation} 
Introducing (\ref{Eq-Gated-Factorization}) into (\ref{Eq-reflection}), we get very simple refraction laws for gated plasmons:
\begin{equation}
 r^G=\frac{q_{pL}^2\sqrt{q_{pR}^2-q_y^2} - q_{pR}^2\sqrt{q_{pL}^2-q_y^2}}{q_{pL}^2\sqrt{q_{pR}^2-q_y^2} +  q_{pR}^2\sqrt{q_{pL}^2-q_y^2}}.
\end{equation} 
The same result could be obtained in a simpler fashion, just by matching the potential and current across the boundary. This approach is correct in the case of strong screening, where electrostatics becomes local. Hence, coincidence of Wiener-Hopf result with wave matching result in the gated case serves as a check for this complex method. On the other hand, for non-gated plasmons the matching approach does not work, and we have to deal with full Wiener-Hopf expressions for reflection and transmission (\ref{Eq-reflection}) and (\ref{Eq-Transmission}).

\begin{figure}
    \includegraphics[width = 0.95\linewidth]{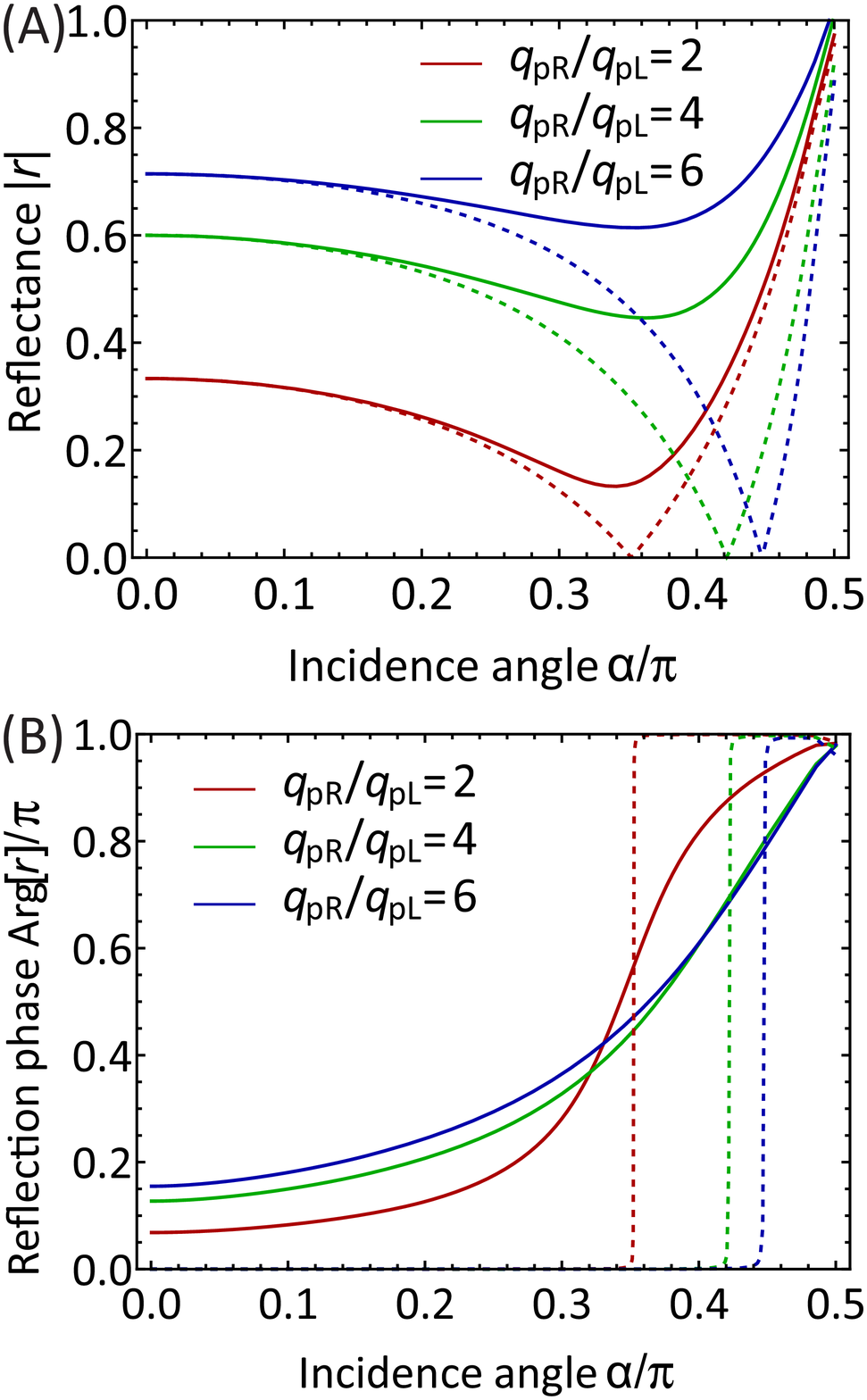}
    \caption{\label{fig2} Computed reflectances $|r|$ and reflection phases ${\rm arg} r$ for a two-dimensional plasmon incident from medium with high conductivity to the medium with low conductivity (${\rm Im}\sigma_L > {\rm Im}\sigma_R$, $q_{p,L} < q_{p, R}$). Solid lines represent the result for non-gated plasmons, while dashed lines correspond to gated 2d plasmons.}
\end{figure}

\begin{figure}
    \includegraphics[width = 0.95\linewidth]{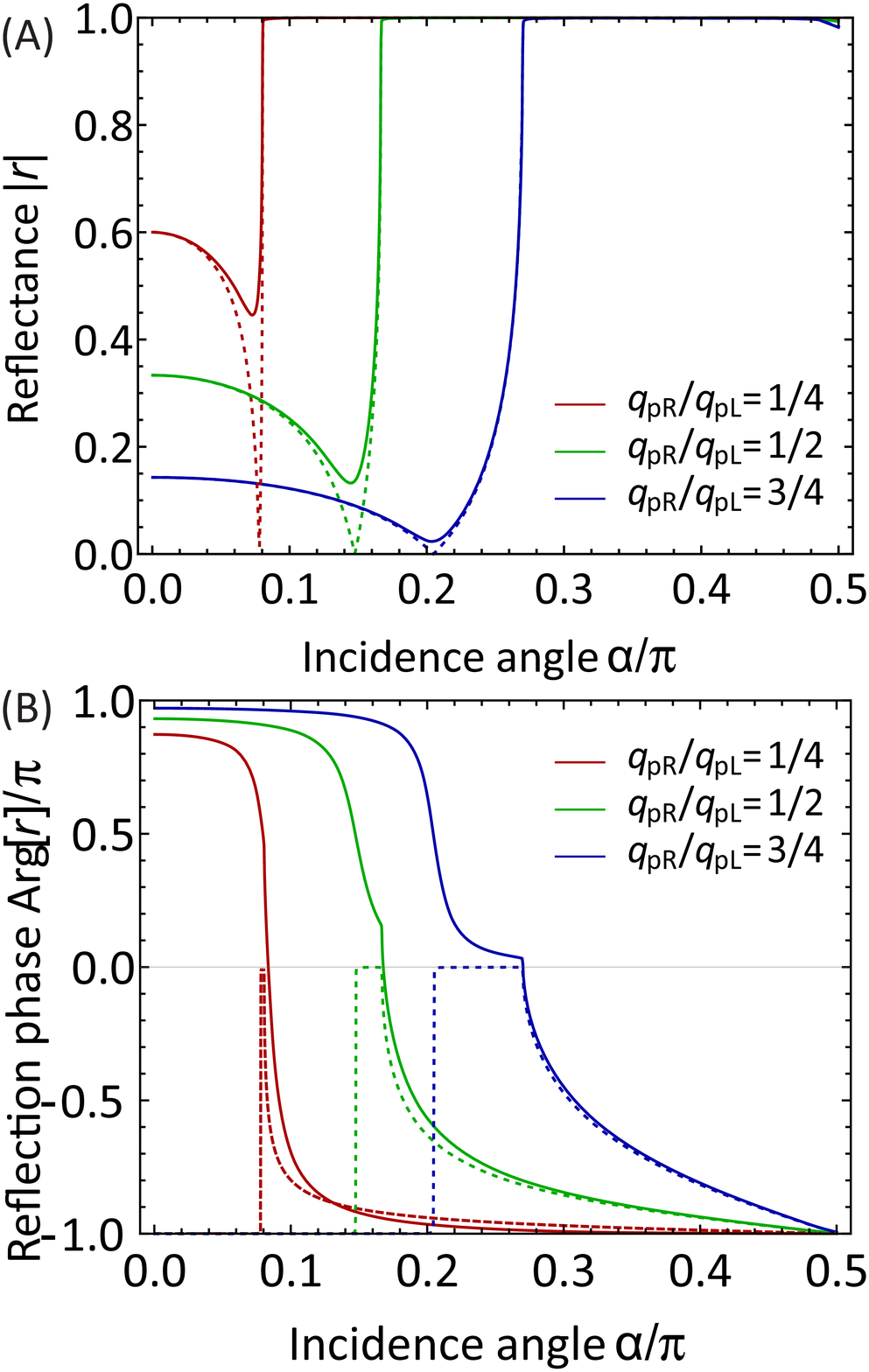}
    \caption{\label{fig3} Computed reflectances $|r|$ and reflection phases ${\rm arg} r$ for a two-dimensional plasmon incident from medium with low conductivity to the medium with high conductivity (${\rm Im}\sigma_L < {\rm Im}\sigma_R$, $q_{p,L} > q_{p, R}$). Solid lines represent the result for non-gated plasmons, while dashed lines correspond to gated 2d plasmons.}
\end{figure}

The computed reflection coefficient, according to Eq.~\ref{Eq-reflection}, is shown in Figs.~\ref{fig2} and \ref{fig3}, for both the absolute value and phase. Expressing the parameters of left and right 2DES sections through the respective absolute values of plasmon wave vector $q_{\rm pL}$ and $q_{\rm pR}$, we can present the result in unified fashion for non-gated and gated plasmons. These are shown with solid and dashed lines, respectively. It is possible to show that absolute reflectance falls down to zero for gated plasmons at the incidence angle $\alpha^*$ satisfying the Brewster-type condition
\begin{equation}
\label{Eq-noreflection1}
    \tan\alpha^* = \frac{q_{pR}}{q_{pL}}.
\end{equation}
Above, we have used the Snell's law $q_{pL}\sin\alpha = q_{pR}\sin\alpha'$ which is valid for arbitrary waves. For non-gated plasmons, the reflectance (\ref{Eq-reflection}) has a dip not reaching zero, but becoming more pronounced for smaller 'contrast' of left and right sections. A similar dip was observed in electromagnetic simulations~\cite{Farajollahi2016}.

It is tempting to associate a dip in transmission with Brewster effect, similarly to conventional optics. In that case, the reflected wave vector should be co-directional with light-induced dipole moment in the medium \# 2, but the dipole intensity in such direction turns to zero. Such explanation does not apply in the case of 2d plasmons, which have electric vector parallel to the propagation direction. The dipole emission intensity is thus always non-zero for directions of 2d plasmon reflection. Finally, the concept of canonical dipole radiation does not apply to 2d plasmons treated in the non-retarded approximation, $c\rightarrow\infty$.

A careful analysis shows that induced dipoles at the boundary of two 2DES sections do not appear at all at the non-reflection angle $\alpha^*$. More precisely, the magnitudes of surface currents ${\bf j}$ in the incident and transmitted wave are fine-tuned to cause no linear charge accumulation at the boundary. As a result, no physical stimulus appears for the reflection. To prove this viewpoint, we rewrite the no-reflection conduction (\ref{Eq-noreflection1}) via conductivity and wave vector
\begin{equation}
\label{Eq-current-matching}
\sigma_L q_x = \sigma_R q'_x.
\end{equation}
This is precisely the condition of current continuity $j_x = -\sigma(x) \varphi'(x)$ between incident and transmitted wave, from which the absence of dynamic charge immediately follows.

Interestingly, the reflection dip for non-gated plasmons never reaches zero. This fact is linked to the scattered fields having the non-propagating spatial structure, evanescent waves. As a result, some charge accumulation in the boundary layer should appear for non-gated plasmons, even despite the fulfillment of current matching condition (\ref{Eq-current-matching}). The latter condition applies only to the plane-wave part of the solution, and does not include evanescent fields.

The second important property of non-gated plasmon reflection, illustrated in Fig.~\ref{fig2} b, is the non-trivial reflection phase shift. It is different from zero and $\pi$, and grows monotonically with increasing the angle of incidence. The case of gated plasmons, shown in the same figure with dashed lines, demonstrates a simpler behavior. The phase shift changes here stepwise between zero and $\pi$ at the non-reflection angle $\alpha^*$. This situation is analogous to the reflection of $p$-polarized waves near the Brewster angle, though the origin of non-reflectance here is completely different.

Minimization of reflection for non-gated plasmons, and its full absence for gated plasmons, persists also for incidence from medium  with low conductivity to the medium with high conductivity. As illustrated in Fig.~\ref{fig3} plotted for this case, the reflectance dip occurs at angles slightly below the angle of total internal reflection $\alpha_{\rm tir}$. Remarkably, the variation of both amplitude and phase of reflection is very abrupt between $\alpha^*$ and $\alpha_{\rm tir}$. These abrupt variations would result in strong Goos-Hanchen shifts for the reflected waves~\cite{Artman_GH_theory}. In our particular 2d setup, Goos-Hanchen shift can be interpreted as excitation of leaky inter-edge plasmons~\cite{IE_Plasmons}. From practical viewpoint, abrupt phase variations can be used for sensing applications, wherein the identified object modifies the properties of 2d conductivity~\cite{Graphene-biosensor}.

The method for obtaining reflection and transmission used here can be extended to include non-local conduction effects, such as electron drift~\cite{DS_Instab,Sydoruk_amplifying} and viscosity~\cite{Cohen}. It can also be applied to scattering of 2d plasmons at the boundary between gated and non-gated domains~\cite{Sydoruk_gated-nongated,Partly-gated,Aizin2012}. The obtained reflection and transmission coefficients can be used as building blocks for design of more complex structures, such as 2d plasmonic crystals with alternating doping~\cite{Muravev-grating,Tombet_amplification,Dyer2013} and 2d plasmonic waveguides. Possible applications of such structures currently lie in the fields of ultra long-wavelength radiation detection, emission, and modulation.

This work was supported by the Russian Science Foundation (Grant No. 21-72-10163).

\bibliography{references}

\begin{thebibliography}{41}%
\makeatletter
\providecommand \@ifxundefined [1]{%
 \@ifx{#1\undefined}
}%
\providecommand \@ifnum [1]{%
 \ifnum #1\expandafter \@firstoftwo
 \else \expandafter \@secondoftwo
 \fi
}%
\providecommand \@ifx [1]{%
 \ifx #1\expandafter \@firstoftwo
 \else \expandafter \@secondoftwo
 \fi
}%
\providecommand \natexlab [1]{#1}%
\providecommand \enquote  [1]{``#1''}%
\providecommand \bibnamefont  [1]{#1}%
\providecommand \bibfnamefont [1]{#1}%
\providecommand \citenamefont [1]{#1}%
\providecommand \href@noop [0]{\@secondoftwo}%
\providecommand \href [0]{\begingroup \@sanitize@url \@href}%
\providecommand \@href[1]{\@@startlink{#1}\@@href}%
\providecommand \@@href[1]{\endgroup#1\@@endlink}%
\providecommand \@sanitize@url [0]{\catcode `\\12\catcode `\$12\catcode
  `\&12\catcode `\#12\catcode `\^12\catcode `\_12\catcode `\%12\relax}%
\providecommand \@@startlink[1]{}%
\providecommand \@@endlink[0]{}%
\providecommand \url  [0]{\begingroup\@sanitize@url \@url }%
\providecommand \@url [1]{\endgroup\@href {#1}{\urlprefix }}%
\providecommand \urlprefix  [0]{URL }%
\providecommand \Eprint [0]{\href }%
\providecommand \doibase [0]{http://dx.doi.org/}%
\providecommand \selectlanguage [0]{\@gobble}%
\providecommand \bibinfo  [0]{\@secondoftwo}%
\providecommand \bibfield  [0]{\@secondoftwo}%
\providecommand \translation [1]{[#1]}%
\providecommand \BibitemOpen [0]{}%
\providecommand \bibitemStop [0]{}%
\providecommand \bibitemNoStop [0]{.\EOS\space}%
\providecommand \EOS [0]{\spacefactor3000\relax}%
\providecommand \BibitemShut  [1]{\csname bibitem#1\endcsname}%
\let\auto@bib@innerbib\@empty
\bibitem [{\citenamefont {Fresnel}(1821)}]{fresnel_augustin_jean_2021_4561712}%
  \BibitemOpen
  \bibfield  {author} {\bibinfo {author} {\bibfnamefont {Augustin-Jean~and}\
  \bibnamefont {Fresnel}},\ }\bibfield  {title} {\enquote {\bibinfo {title} {On
  the calculation of the tints that polarization develops in crystalline
  plates},}\ }\href {\doibase 10.5281/zenodo.4561712} {\bibfield  {journal}
  {\bibinfo  {journal} {Annales de Chimie et de Physique}\ }\textbf {\bibinfo
  {volume} {17}},\ \bibinfo {pages} {102–111} (\bibinfo {year}
  {1821})}\BibitemShut {NoStop}%
\bibitem [{\citenamefont {Landau}\ and\ \citenamefont
  {Lifshitz}(2013)}]{landau2013fluid}%
  \BibitemOpen
  \bibfield  {author} {\bibinfo {author} {\bibfnamefont {Lev~Davidovich}\
  \bibnamefont {Landau}}\ and\ \bibinfo {author} {\bibfnamefont
  {Evgenii~Mikhailovich}\ \bibnamefont {Lifshitz}},\ }\href@noop {} {\emph
  {\bibinfo {title} {Fluid Mechanics. Course of Theoretical Physics}}},\
  Vol.~\bibinfo {volume} {6}\ (\bibinfo  {publisher} {Elsevier},\ \bibinfo
  {year} {2013})\BibitemShut {NoStop}%
\bibitem [{\citenamefont {Stern}(1967)}]{Stern1967}%
  \BibitemOpen
  \bibfield  {author} {\bibinfo {author} {\bibfnamefont {Frank}\ \bibnamefont
  {Stern}},\ }\bibfield  {title} {\enquote {\bibinfo {title} {{Polarizability
  of a Two-Dimensional Electron Gas}},}\ }\href {\doibase
  10.1103/PhysRevLett.18.546} {\bibfield  {journal} {\bibinfo  {journal}
  {Physical Review Letters}\ }\textbf {\bibinfo {volume} {18}},\ \bibinfo
  {pages} {546--548} (\bibinfo {year} {1967})}\BibitemShut {NoStop}%
\bibitem [{\citenamefont {Allen}\ \emph {et~al.}(1977)\citenamefont {Allen},
  \citenamefont {Tsui},\ and\ \citenamefont {Logan}}]{Allen1977}%
  \BibitemOpen
  \bibfield  {author} {\bibinfo {author} {\bibfnamefont {S.~J.}\ \bibnamefont
  {Allen}}, \bibinfo {author} {\bibfnamefont {D.~C.}\ \bibnamefont {Tsui}}, \
  and\ \bibinfo {author} {\bibfnamefont {R.~A.}\ \bibnamefont {Logan}},\
  }\bibfield  {title} {\enquote {\bibinfo {title} {{Observation of the
  two-dimensional plasmon in silicon inversion layers}},}\ }\href {\doibase
  10.1103/PhysRevLett.38.980} {\bibfield  {journal} {\bibinfo  {journal}
  {Physical Review Letters}\ }\textbf {\bibinfo {volume} {38}},\ \bibinfo
  {pages} {980--983} (\bibinfo {year} {1977})}\BibitemShut {NoStop}%
\bibitem [{\citenamefont {Chaplik}(1972)}]{Chaplik1972}%
  \BibitemOpen
  \bibfield  {author} {\bibinfo {author} {\bibfnamefont {A.V.}\ \bibnamefont
  {Chaplik}},\ }\bibfield  {title} {\enquote {\bibinfo {title} {{Possible
  Crystallization of Charge Carriers in Low-density Inversion Layers}},}\
  }\href@noop {} {\bibfield  {journal} {\bibinfo  {journal} {Sov. Phys. JETP}\
  }\textbf {\bibinfo {volume} {35}},\ \bibinfo {pages} {395} (\bibinfo {year}
  {1972})}\BibitemShut {NoStop}%
\bibitem [{\citenamefont {Iranzo}\ \emph {et~al.}(2018)\citenamefont {Iranzo},
  \citenamefont {Nanot}, \citenamefont {Dias}, \citenamefont {Epstein},
  \citenamefont {Peng}, \citenamefont {Efetov}, \citenamefont {Lundeberg},
  \citenamefont {Parret}, \citenamefont {Osmond}, \citenamefont {Hong},
  \citenamefont {Kong}, \citenamefont {Englund}, \citenamefont {Peres},\ and\
  \citenamefont {Koppens}}]{Iranzo2018}%
  \BibitemOpen
  \bibfield  {author} {\bibinfo {author} {\bibfnamefont {David~Alcaraz}\
  \bibnamefont {Iranzo}}, \bibinfo {author} {\bibfnamefont {S{\'{e}}bastien}\
  \bibnamefont {Nanot}}, \bibinfo {author} {\bibfnamefont {Eduardo~J.C.}\
  \bibnamefont {Dias}}, \bibinfo {author} {\bibfnamefont {Itai}\ \bibnamefont
  {Epstein}}, \bibinfo {author} {\bibfnamefont {Cheng}\ \bibnamefont {Peng}},
  \bibinfo {author} {\bibfnamefont {Dmitri~K.}\ \bibnamefont {Efetov}},
  \bibinfo {author} {\bibfnamefont {Mark~B.}\ \bibnamefont {Lundeberg}},
  \bibinfo {author} {\bibfnamefont {Romain}\ \bibnamefont {Parret}}, \bibinfo
  {author} {\bibfnamefont {Johann}\ \bibnamefont {Osmond}}, \bibinfo {author}
  {\bibfnamefont {Jin~Yong}\ \bibnamefont {Hong}}, \bibinfo {author}
  {\bibfnamefont {Jing}\ \bibnamefont {Kong}}, \bibinfo {author} {\bibfnamefont
  {Dirk~R.}\ \bibnamefont {Englund}}, \bibinfo {author} {\bibfnamefont
  {Nuno~M.R.}\ \bibnamefont {Peres}}, \ and\ \bibinfo {author} {\bibfnamefont
  {Frank~H.L.}\ \bibnamefont {Koppens}},\ }\bibfield  {title} {\enquote
  {\bibinfo {title} {{Probing the ultimate plasmon confinement limits with a
  van der Waals heterostructure}},}\ }\href {\doibase 10.1126/science.aar8438}
  {\bibfield  {journal} {\bibinfo  {journal} {Science}\ }\textbf {\bibinfo
  {volume} {360}},\ \bibinfo {pages} {291--295} (\bibinfo {year} {2018})},\
  \Eprint {http://arxiv.org/abs/1804.01061} {1804.01061} \BibitemShut {NoStop}%
\bibitem [{\citenamefont {Muravev}\ \emph {et~al.}(2016)\citenamefont
  {Muravev}, \citenamefont {Fortunatov}, \citenamefont {Dremin},\ and\
  \citenamefont {Kukushkin}}]{Muravev2016a}%
  \BibitemOpen
  \bibfield  {author} {\bibinfo {author} {\bibfnamefont {V.~M.}\ \bibnamefont
  {Muravev}}, \bibinfo {author} {\bibfnamefont {A.~A.}\ \bibnamefont
  {Fortunatov}}, \bibinfo {author} {\bibfnamefont {A.~A.}\ \bibnamefont
  {Dremin}}, \ and\ \bibinfo {author} {\bibfnamefont {I.~V.}\ \bibnamefont
  {Kukushkin}},\ }\bibfield  {title} {\enquote {\bibinfo {title} {{Plasmonic
  interferometer for spectroscopy of microwave radiation}},}\ }\href {\doibase
  10.1134/S0021364016060084} {\bibfield  {journal} {\bibinfo  {journal} {JETP
  Letters}\ }\textbf {\bibinfo {volume} {103}},\ \bibinfo {pages} {380--384}
  (\bibinfo {year} {2016})}\BibitemShut {NoStop}%
\bibitem [{\citenamefont {Knap}\ \emph {et~al.}(2002)\citenamefont {Knap},
  \citenamefont {Deng}, \citenamefont {Rumyantsev}, \citenamefont {L{\"{u}}},
  \citenamefont {Shur}, \citenamefont {Saylor},\ and\ \citenamefont
  {Brunel}}]{Knap2002}%
  \BibitemOpen
  \bibfield  {author} {\bibinfo {author} {\bibfnamefont {W.}~\bibnamefont
  {Knap}}, \bibinfo {author} {\bibfnamefont {Y.}~\bibnamefont {Deng}}, \bibinfo
  {author} {\bibfnamefont {S.}~\bibnamefont {Rumyantsev}}, \bibinfo {author}
  {\bibfnamefont {J.-Q.}\ \bibnamefont {L{\"{u}}}}, \bibinfo {author}
  {\bibfnamefont {M.~S.}\ \bibnamefont {Shur}}, \bibinfo {author}
  {\bibfnamefont {C.~A.}\ \bibnamefont {Saylor}}, \ and\ \bibinfo {author}
  {\bibfnamefont {L.~C.}\ \bibnamefont {Brunel}},\ }\bibfield  {title}
  {\enquote {\bibinfo {title} {{Resonant detection of subterahertz radiation by
  plasma waves in a submicron field-effect transistor}},}\ }\href {\doibase
  10.1063/1.1473685} {\bibfield  {journal} {\bibinfo  {journal} {Applied
  Physics Letters}\ }\textbf {\bibinfo {volume} {80}},\ \bibinfo {pages}
  {3433--3435} (\bibinfo {year} {2002})}\BibitemShut {NoStop}%
\bibitem [{\citenamefont {Bandurin}\ \emph {et~al.}(2018)\citenamefont
  {Bandurin}, \citenamefont {Svintsov}, \citenamefont {Gayduchenko},
  \citenamefont {Xu}, \citenamefont {Principi}, \citenamefont {Moskotin},
  \citenamefont {Tretyakov}, \citenamefont {Yagodkin}, \citenamefont {Zhukov},
  \citenamefont {Taniguchi}, \citenamefont {Watanabe}, \citenamefont
  {Grigorieva}, \citenamefont {Polini}, \citenamefont {Goltsman}, \citenamefont
  {Geim},\ and\ \citenamefont {Fedorov}}]{Bandurin2018d}%
  \BibitemOpen
  \bibfield  {author} {\bibinfo {author} {\bibfnamefont {Denis~A.}\
  \bibnamefont {Bandurin}}, \bibinfo {author} {\bibfnamefont {Dmitry}\
  \bibnamefont {Svintsov}}, \bibinfo {author} {\bibfnamefont {Igor}\
  \bibnamefont {Gayduchenko}}, \bibinfo {author} {\bibfnamefont {Shuigang~G.}\
  \bibnamefont {Xu}}, \bibinfo {author} {\bibfnamefont {Alessandro}\
  \bibnamefont {Principi}}, \bibinfo {author} {\bibfnamefont {Maxim}\
  \bibnamefont {Moskotin}}, \bibinfo {author} {\bibfnamefont {Ivan}\
  \bibnamefont {Tretyakov}}, \bibinfo {author} {\bibfnamefont {Denis}\
  \bibnamefont {Yagodkin}}, \bibinfo {author} {\bibfnamefont {Sergey}\
  \bibnamefont {Zhukov}}, \bibinfo {author} {\bibfnamefont {Takashi}\
  \bibnamefont {Taniguchi}}, \bibinfo {author} {\bibfnamefont {Kenji}\
  \bibnamefont {Watanabe}}, \bibinfo {author} {\bibfnamefont {Irina~V.}\
  \bibnamefont {Grigorieva}}, \bibinfo {author} {\bibfnamefont {Marco}\
  \bibnamefont {Polini}}, \bibinfo {author} {\bibfnamefont {Gregory~N.}\
  \bibnamefont {Goltsman}}, \bibinfo {author} {\bibfnamefont {Andre~K.}\
  \bibnamefont {Geim}}, \ and\ \bibinfo {author} {\bibfnamefont {Georgy}\
  \bibnamefont {Fedorov}},\ }\bibfield  {title} {\enquote {\bibinfo {title}
  {{Resonant terahertz detection using graphene plasmons}},}\ }\href {\doibase
  10.1038/s41467-018-07848-w} {\bibfield  {journal} {\bibinfo  {journal}
  {Nature Communications}\ }\textbf {\bibinfo {volume} {9}},\ \bibinfo {pages}
  {5392} (\bibinfo {year} {2018})},\ \Eprint {http://arxiv.org/abs/1807.04703}
  {1807.04703} \BibitemShut {NoStop}%
\bibitem [{\citenamefont {{El Fatimy}}\ \emph {et~al.}(2010)\citenamefont {{El
  Fatimy}}, \citenamefont {Dyakonova}, \citenamefont {Meziani}, \citenamefont
  {Otsuji}, \citenamefont {Knap}, \citenamefont {Vandenbrouk}, \citenamefont
  {Madjour}, \citenamefont {Th{\'{e}}ron}, \citenamefont {Gaquiere},
  \citenamefont {Poisson}, \citenamefont {Delage}, \citenamefont {Prystawko},\
  and\ \citenamefont {Skierbiszewski}}]{ElFatimy2010}%
  \BibitemOpen
  \bibfield  {author} {\bibinfo {author} {\bibfnamefont {A.}~\bibnamefont {{El
  Fatimy}}}, \bibinfo {author} {\bibfnamefont {N.}~\bibnamefont {Dyakonova}},
  \bibinfo {author} {\bibfnamefont {Y.}~\bibnamefont {Meziani}}, \bibinfo
  {author} {\bibfnamefont {T.}~\bibnamefont {Otsuji}}, \bibinfo {author}
  {\bibfnamefont {W.}~\bibnamefont {Knap}}, \bibinfo {author} {\bibfnamefont
  {S.}~\bibnamefont {Vandenbrouk}}, \bibinfo {author} {\bibfnamefont
  {K.}~\bibnamefont {Madjour}}, \bibinfo {author} {\bibfnamefont
  {D.}~\bibnamefont {Th{\'{e}}ron}}, \bibinfo {author} {\bibfnamefont
  {C.}~\bibnamefont {Gaquiere}}, \bibinfo {author} {\bibfnamefont {M.~A.}\
  \bibnamefont {Poisson}}, \bibinfo {author} {\bibfnamefont {S.}~\bibnamefont
  {Delage}}, \bibinfo {author} {\bibfnamefont {P.}~\bibnamefont {Prystawko}}, \
  and\ \bibinfo {author} {\bibfnamefont {C.}~\bibnamefont {Skierbiszewski}},\
  }\bibfield  {title} {\enquote {\bibinfo {title} {{AlGaN/GaN high electron
  mobility transistors as a voltage-tunable room temperature terahertz
  sources}},}\ }\href {\doibase 10.1063/1.3291101} {\bibfield  {journal}
  {\bibinfo  {journal} {Journal of Applied Physics}\ }\textbf {\bibinfo
  {volume} {107}},\ \bibinfo {pages} {024504} (\bibinfo {year}
  {2010})}\BibitemShut {NoStop}%
\bibitem [{\citenamefont {Boubanga-Tombet}\ \emph {et~al.}(2020)\citenamefont
  {Boubanga-Tombet}, \citenamefont {Knap}, \citenamefont {Yadav}, \citenamefont
  {Satou}, \citenamefont {But}, \citenamefont {Popov}, \citenamefont
  {Gorbenko}, \citenamefont {Kachorovskii},\ and\ \citenamefont
  {Otsuji}}]{Tombet_amplification}%
  \BibitemOpen
  \bibfield  {author} {\bibinfo {author} {\bibfnamefont {Stephane}\
  \bibnamefont {Boubanga-Tombet}}, \bibinfo {author} {\bibfnamefont {Wojciech}\
  \bibnamefont {Knap}}, \bibinfo {author} {\bibfnamefont {Deepika}\
  \bibnamefont {Yadav}}, \bibinfo {author} {\bibfnamefont {Akira}\ \bibnamefont
  {Satou}}, \bibinfo {author} {\bibfnamefont {Dmytro~B.}\ \bibnamefont {But}},
  \bibinfo {author} {\bibfnamefont {Vyacheslav~V.}\ \bibnamefont {Popov}},
  \bibinfo {author} {\bibfnamefont {Ilya~V.}\ \bibnamefont {Gorbenko}},
  \bibinfo {author} {\bibfnamefont {Valentin}\ \bibnamefont {Kachorovskii}}, \
  and\ \bibinfo {author} {\bibfnamefont {Taiichi}\ \bibnamefont {Otsuji}},\
  }\bibfield  {title} {\enquote {\bibinfo {title} {Room-temperature
  amplification of terahertz radiation by grating-gate graphene structures},}\
  }\href {\doibase 10.1103/PhysRevX.10.031004} {\bibfield  {journal} {\bibinfo
  {journal} {Phys. Rev. X}\ }\textbf {\bibinfo {volume} {10}},\ \bibinfo
  {pages} {031004} (\bibinfo {year} {2020})}\BibitemShut {NoStop}%
\bibitem [{\citenamefont {Orgiu}\ \emph {et~al.}(2015)\citenamefont {Orgiu},
  \citenamefont {George}, \citenamefont {Hutchison}, \citenamefont {Devaux},
  \citenamefont {Dayen}, \citenamefont {Doudin}, \citenamefont {Stellacci},
  \citenamefont {Genet}, \citenamefont {Schachenmayer}, \citenamefont {Genes},
  \citenamefont {Pupillo}, \citenamefont {Samor{\`{i}}},\ and\ \citenamefont
  {Ebbesen}}]{Orgiu2015}%
  \BibitemOpen
  \bibfield  {author} {\bibinfo {author} {\bibfnamefont {E.}~\bibnamefont
  {Orgiu}}, \bibinfo {author} {\bibfnamefont {J.}~\bibnamefont {George}},
  \bibinfo {author} {\bibfnamefont {J.~A.}\ \bibnamefont {Hutchison}}, \bibinfo
  {author} {\bibfnamefont {E.}~\bibnamefont {Devaux}}, \bibinfo {author}
  {\bibfnamefont {J.~F.}\ \bibnamefont {Dayen}}, \bibinfo {author}
  {\bibfnamefont {B.}~\bibnamefont {Doudin}}, \bibinfo {author} {\bibfnamefont
  {F.}~\bibnamefont {Stellacci}}, \bibinfo {author} {\bibfnamefont
  {C.}~\bibnamefont {Genet}}, \bibinfo {author} {\bibfnamefont
  {J.}~\bibnamefont {Schachenmayer}}, \bibinfo {author} {\bibfnamefont
  {C.}~\bibnamefont {Genes}}, \bibinfo {author} {\bibfnamefont
  {G.}~\bibnamefont {Pupillo}}, \bibinfo {author} {\bibfnamefont
  {P.}~\bibnamefont {Samor{\`{i}}}}, \ and\ \bibinfo {author} {\bibfnamefont
  {T.~W.}\ \bibnamefont {Ebbesen}},\ }\bibfield  {title} {\enquote {\bibinfo
  {title} {{Conductivity in organic semiconductors hybridized with the vacuum
  field}},}\ }\href {\doibase 10.1038/nmat4392} {\bibfield  {journal} {\bibinfo
   {journal} {Nature Materials}\ }\textbf {\bibinfo {volume} {14}},\ \bibinfo
  {pages} {1123--1129} (\bibinfo {year} {2015})},\ \Eprint
  {http://arxiv.org/abs/1409.1900} {1409.1900} \BibitemShut {NoStop}%
\bibitem [{\citenamefont {Jiang}\ \emph {et~al.}(2018)\citenamefont {Jiang},
  \citenamefont {Mele},\ and\ \citenamefont {Fogler}}]{Fogler2018}%
  \BibitemOpen
  \bibfield  {author} {\bibinfo {author} {\bibfnamefont {Bor-Yuan}\
  \bibnamefont {Jiang}}, \bibinfo {author} {\bibfnamefont {Eugene~J.}\
  \bibnamefont {Mele}}, \ and\ \bibinfo {author} {\bibfnamefont {Michael~M.}\
  \bibnamefont {Fogler}},\ }\bibfield  {title} {\enquote {\bibinfo {title}
  {{Theory of plasmon reflection by a 1D junction}},}\ }\href {\doibase
  10.1364/oe.26.017209} {\bibfield  {journal} {\bibinfo  {journal} {Optics
  Express}\ }\textbf {\bibinfo {volume} {26}},\ \bibinfo {pages} {17209}
  (\bibinfo {year} {2018})},\ \Eprint {http://arxiv.org/abs/1804.05256}
  {1804.05256} \BibitemShut {NoStop}%
\bibitem [{\citenamefont {Siaber}\ \emph {et~al.}(2019)\citenamefont {Siaber},
  \citenamefont {Zonetti},\ and\ \citenamefont {Sydoruk}}]{Siaber2019}%
  \BibitemOpen
  \bibfield  {author} {\bibinfo {author} {\bibfnamefont {S.}~\bibnamefont
  {Siaber}}, \bibinfo {author} {\bibfnamefont {S.}~\bibnamefont {Zonetti}}, \
  and\ \bibinfo {author} {\bibfnamefont {O.}~\bibnamefont {Sydoruk}},\
  }\bibfield  {title} {\enquote {\bibinfo {title} {{Junctions between
  two-dimensional plasmonic waveguides in the presence of retardation}},}\
  }\href {\doibase 10.1088/2040-8986/ab4056} {\bibfield  {journal} {\bibinfo
  {journal} {Journal of Optics}\ }\textbf {\bibinfo {volume} {21}},\ \bibinfo
  {pages} {105002} (\bibinfo {year} {2019})}\BibitemShut {NoStop}%
\bibitem [{\citenamefont {Semenenko}\ \emph {et~al.}(2020)\citenamefont
  {Semenenko}, \citenamefont {Liu},\ and\ \citenamefont
  {Perebeinos}}]{Semenenko_scattering}%
  \BibitemOpen
  \bibfield  {author} {\bibinfo {author} {\bibfnamefont {Vyacheslav}\
  \bibnamefont {Semenenko}}, \bibinfo {author} {\bibfnamefont {Mengkun}\
  \bibnamefont {Liu}}, \ and\ \bibinfo {author} {\bibfnamefont {Vasili}\
  \bibnamefont {Perebeinos}},\ }\bibfield  {title} {\enquote {\bibinfo {title}
  {Scattering of quasistatic plasmons from one-dimensional junctions of
  graphene: Transfer matrices, fresnel relations, and nonlocality},}\ }\href
  {\doibase 10.1103/PhysRevApplied.14.024049} {\bibfield  {journal} {\bibinfo
  {journal} {Phys. Rev. Appl.}\ }\textbf {\bibinfo {volume} {14}},\ \bibinfo
  {pages} {024049} (\bibinfo {year} {2020})}\BibitemShut {NoStop}%
\bibitem [{\citenamefont {Chaves}\ \emph {et~al.}(2018)\citenamefont {Chaves},
  \citenamefont {Amorim}, \citenamefont {Bludov}, \citenamefont {Goncalves},\
  and\ \citenamefont {Peres}}]{Peres_PRB_Scattering}%
  \BibitemOpen
  \bibfield  {author} {\bibinfo {author} {\bibfnamefont {A.~J.}\ \bibnamefont
  {Chaves}}, \bibinfo {author} {\bibfnamefont {B.}~\bibnamefont {Amorim}},
  \bibinfo {author} {\bibfnamefont {Yu.~V.}\ \bibnamefont {Bludov}}, \bibinfo
  {author} {\bibfnamefont {P.~A.~D.}\ \bibnamefont {Goncalves}}, \ and\
  \bibinfo {author} {\bibfnamefont {N.~M.~R.}\ \bibnamefont {Peres}},\
  }\bibfield  {title} {\enquote {\bibinfo {title} {Scattering of graphene
  plasmons at abrupt interfaces: An analytic and numeric study},}\ }\href
  {\doibase 10.1103/PhysRevB.97.035434} {\bibfield  {journal} {\bibinfo
  {journal} {Phys. Rev. B}\ }\textbf {\bibinfo {volume} {97}},\ \bibinfo
  {pages} {035434} (\bibinfo {year} {2018})}\BibitemShut {NoStop}%
\bibitem [{\citenamefont {Farajollahi}\ \emph {et~al.}(2016)\citenamefont
  {Farajollahi}, \citenamefont {Rejaei},\ and\ \citenamefont
  {Khavasi}}]{Farajollahi2016}%
  \BibitemOpen
  \bibfield  {author} {\bibinfo {author} {\bibfnamefont {Saeed}\ \bibnamefont
  {Farajollahi}}, \bibinfo {author} {\bibfnamefont {Behzad}\ \bibnamefont
  {Rejaei}}, \ and\ \bibinfo {author} {\bibfnamefont {Amin}\ \bibnamefont
  {Khavasi}},\ }\bibfield  {title} {\enquote {\bibinfo {title} {{Reflection and
  transmission of obliquely incident graphene plasmons by discontinuities in
  surface conductivity: observation of the Brewster-like effect}},}\ }\href
  {\doibase 10.1088/2040-8978/18/7/075005} {\bibfield  {journal} {\bibinfo
  {journal} {Journal of Optics}\ }\textbf {\bibinfo {volume} {18}},\ \bibinfo
  {pages} {075005} (\bibinfo {year} {2016})}\BibitemShut {NoStop}%
\bibitem [{\citenamefont {Senior}(1952)}]{Senior_WienerHopf}%
  \BibitemOpen
  \bibfield  {author} {\bibinfo {author} {\bibfnamefont {T.~B.A.}\ \bibnamefont
  {Senior}},\ }\bibfield  {title} {\enquote {\bibinfo {title} {{Diffraction by
  a semi-infinite metallic sheet}},}\ }\href {\doibase 10.1098/rspa.1952.0137}
  {\bibfield  {journal} {\bibinfo  {journal} {Proceedings of the Royal Society
  of London. Series A. Mathematical and Physical Sciences}\ }\textbf {\bibinfo
  {volume} {213}},\ \bibinfo {pages} {436--458} (\bibinfo {year}
  {1952})}\BibitemShut {NoStop}%
\bibitem [{\citenamefont {Nussenzveig}\ and\ \citenamefont
  {Lighthill}(1959)}]{Waveguide_diffraction}%
  \BibitemOpen
  \bibfield  {author} {\bibinfo {author} {\bibfnamefont {H.~M.}\ \bibnamefont
  {Nussenzveig}}\ and\ \bibinfo {author} {\bibfnamefont {Michael~James}\
  \bibnamefont {Lighthill}},\ }\bibfield  {title} {\enquote {\bibinfo {title}
  {Solution of a diffraction problem - solution of a diffraction problem. i the
  wide double wedge},}\ }\href {\doibase 10.1098/rsta.1959.0012} {\bibfield
  {journal} {\bibinfo  {journal} {Philosophical Transactions of the Royal
  Society of London. Series A, Mathematical and Physical Sciences}\ }\textbf
  {\bibinfo {volume} {252}},\ \bibinfo {pages} {1--30} (\bibinfo {year}
  {1959})}\BibitemShut {NoStop}%
\bibitem [{\citenamefont {Kay}(1959)}]{Kay1959}%
  \BibitemOpen
  \bibfield  {author} {\bibinfo {author} {\bibfnamefont {A.}~\bibnamefont
  {Kay}},\ }\bibfield  {title} {\enquote {\bibinfo {title} {{Scattering of a
  surface wave by a discontinuity in reactance}},}\ }\href {\doibase
  10.1109/TAP.1959.1144635} {\bibfield  {journal} {\bibinfo  {journal} {IRE
  Transactions on Antennas and Propagation}\ }\textbf {\bibinfo {volume} {7}},\
  \bibinfo {pages} {22--31} (\bibinfo {year} {1959})}\BibitemShut {NoStop}%
\bibitem [{\citenamefont {Rejaei}\ and\ \citenamefont
  {Khavasi}(2015)}]{Rejaei2015}%
  \BibitemOpen
  \bibfield  {author} {\bibinfo {author} {\bibfnamefont {Behzad}\ \bibnamefont
  {Rejaei}}\ and\ \bibinfo {author} {\bibfnamefont {Amin}\ \bibnamefont
  {Khavasi}},\ }\bibfield  {title} {\enquote {\bibinfo {title} {{Scattering of
  surface plasmons on graphene by a discontinuity in surface conductivity}},}\
  }\href {\doibase 10.1088/2040-8978/17/7/075002} {\bibfield  {journal}
  {\bibinfo  {journal} {Journal of Optics (United Kingdom)}\ }\textbf {\bibinfo
  {volume} {17}},\ \bibinfo {pages} {75002} (\bibinfo {year}
  {2015})}\BibitemShut {NoStop}%
\bibitem [{\citenamefont {Daniele}\ \emph {et~al.}(2014)\citenamefont
  {Daniele}, \citenamefont {Zich} \emph {et~al.}}]{daniele2014wiener}%
  \BibitemOpen
  \bibfield  {author} {\bibinfo {author} {\bibfnamefont {Vito~G}\ \bibnamefont
  {Daniele}}, \bibinfo {author} {\bibfnamefont {Rodolfo}\ \bibnamefont {Zich}},
   \emph {et~al.},\ }\href@noop {} {\emph {\bibinfo {title} {The Wiener-Hopf
  method in electromagnetics}}}\ (\bibinfo  {publisher} {SciTech Publishing
  Incorporated},\ \bibinfo {year} {2014})\BibitemShut {NoStop}%
\bibitem [{\citenamefont {Zabolotnykh}\ and\ \citenamefont
  {Volkov}(2016)}]{zabolotnykh2016edge}%
  \BibitemOpen
  \bibfield  {author} {\bibinfo {author} {\bibfnamefont
  {Andrei~Aleksandrovich}\ \bibnamefont {Zabolotnykh}}\ and\ \bibinfo {author}
  {\bibfnamefont {VA}~\bibnamefont {Volkov}},\ }\bibfield  {title} {\enquote
  {\bibinfo {title} {Edge plasmon polaritons on a half-plane},}\ }\href
  {\doibase 10.1134/S0021364016180144} {\bibfield  {journal} {\bibinfo
  {journal} {JETP letters}\ }\textbf {\bibinfo {volume} {104}},\ \bibinfo
  {pages} {411--416} (\bibinfo {year} {2016})}\BibitemShut {NoStop}%
\bibitem [{\citenamefont {Volkov}\ and\ \citenamefont
  {Mikhailov}(1988)}]{volkov1988edge}%
  \BibitemOpen
  \bibfield  {author} {\bibinfo {author} {\bibfnamefont {VA}~\bibnamefont
  {Volkov}}\ and\ \bibinfo {author} {\bibfnamefont {Sergey~A}\ \bibnamefont
  {Mikhailov}},\ }\bibfield  {title} {\enquote {\bibinfo {title} {Edge
  magnetoplasmons: low frequency weakly damped excitations in inhomogeneous
  two-dimensional electron systems},}\ }\href@noop {} {\bibfield  {journal}
  {\bibinfo  {journal} {Sov. Phys. JETP}\ }\textbf {\bibinfo {volume} {67}},\
  \bibinfo {pages} {1639--1653} (\bibinfo {year} {1988})}\BibitemShut {NoStop}%
\bibitem [{\citenamefont {Principi}\ \emph {et~al.}(2016)\citenamefont
  {Principi}, \citenamefont {Katsnelson},\ and\ \citenamefont
  {Vignale}}]{2Component_emps}%
  \BibitemOpen
  \bibfield  {author} {\bibinfo {author} {\bibfnamefont {Alessandro}\
  \bibnamefont {Principi}}, \bibinfo {author} {\bibfnamefont {Mikhail~I.}\
  \bibnamefont {Katsnelson}}, \ and\ \bibinfo {author} {\bibfnamefont
  {Giovanni}\ \bibnamefont {Vignale}},\ }\bibfield  {title} {\enquote {\bibinfo
  {title} {{Edge Plasmons in Two-Component Electron Liquids in the Presence of
  Pseudomagnetic Fields}},}\ }\href {\doibase 10.1103/PhysRevLett.117.196803}
  {\bibfield  {journal} {\bibinfo  {journal} {Physical Review Letters}\
  }\textbf {\bibinfo {volume} {117}},\ \bibinfo {pages} {196803} (\bibinfo
  {year} {2016})}\BibitemShut {NoStop}%
\bibitem [{\citenamefont {Petrov}(2021)}]{Petrov_IEBP}%
  \BibitemOpen
  \bibfield  {author} {\bibinfo {author} {\bibfnamefont {Aleksandr~S.}\
  \bibnamefont {Petrov}},\ }\bibfield  {title} {\enquote {\bibinfo {title}
  {Plasmonic excitation for a tunable transmitter without magnetic field immune
  to backscattering},}\ }\href {\doibase 10.1103/PhysRevB.104.L241407}
  {\bibfield  {journal} {\bibinfo  {journal} {Phys. Rev. B}\ }\textbf {\bibinfo
  {volume} {104}},\ \bibinfo {pages} {L241407} (\bibinfo {year}
  {2021})}\BibitemShut {NoStop}%
\bibitem [{\citenamefont {Sokolik}\ \emph {et~al.}(2021)\citenamefont
  {Sokolik}, \citenamefont {Kotov},\ and\ \citenamefont
  {Lozovik}}]{Lozovik-Sokolik}%
  \BibitemOpen
  \bibfield  {author} {\bibinfo {author} {\bibfnamefont {Alexey~A.}\
  \bibnamefont {Sokolik}}, \bibinfo {author} {\bibfnamefont {Oleg~V.}\
  \bibnamefont {Kotov}}, \ and\ \bibinfo {author} {\bibfnamefont {Yurii~E.}\
  \bibnamefont {Lozovik}},\ }\bibfield  {title} {\enquote {\bibinfo {title}
  {Plasmonic modes at inclined edges of anisotropic two-dimensional
  materials},}\ }\href {\doibase 10.1103/PhysRevB.103.155402} {\bibfield
  {journal} {\bibinfo  {journal} {Phys. Rev. B}\ }\textbf {\bibinfo {volume}
  {103}},\ \bibinfo {pages} {155402} (\bibinfo {year} {2021})}\BibitemShut
  {NoStop}%
\bibitem [{\citenamefont {and and}(2014)}]{Zhang_2014}%
  \BibitemOpen
  \bibfield  {author} {\bibinfo {author} {\bibnamefont {and and}},\ }\bibfield
  {title} {\enquote {\bibinfo {title} {Excitation of propagating plasmons in
  semi-infinite graphene layer by free space photons},}\ }\href {\doibase
  10.1088/0253-6102/61/6/14} {\bibfield  {journal} {\bibinfo  {journal}
  {Communications in Theoretical Physics}\ }\textbf {\bibinfo {volume} {61}},\
  \bibinfo {pages} {751} (\bibinfo {year} {2014})}\BibitemShut {NoStop}%
\bibitem [{\citenamefont {Margetis}\ \emph {et~al.}(2017)\citenamefont
  {Margetis}, \citenamefont {Maier},\ and\ \citenamefont
  {Luskin}}]{Margetis_edge_diffraction}%
  \BibitemOpen
  \bibfield  {author} {\bibinfo {author} {\bibfnamefont {Dionisios}\
  \bibnamefont {Margetis}}, \bibinfo {author} {\bibfnamefont {Matthias}\
  \bibnamefont {Maier}}, \ and\ \bibinfo {author} {\bibfnamefont {Mitchell}\
  \bibnamefont {Luskin}},\ }\bibfield  {title} {\enquote {\bibinfo {title} {On
  the wiener–hopf method for surface plasmons: Diffraction from semiinfinite
  metamaterial sheet},}\ }\href {\doibase https://doi.org/10.1111/sapm.12180}
  {\bibfield  {journal} {\bibinfo  {journal} {Studies in Applied Mathematics}\
  }\textbf {\bibinfo {volume} {139}},\ \bibinfo {pages} {599--625} (\bibinfo
  {year} {2017})}\BibitemShut {NoStop}%
\bibitem [{\citenamefont {Nikulin}\ \emph {et~al.}(2021)\citenamefont
  {Nikulin}, \citenamefont {Mylnikov}, \citenamefont {Bandurin},\ and\
  \citenamefont {Svintsov}}]{Nikulin2021}%
  \BibitemOpen
  \bibfield  {author} {\bibinfo {author} {\bibfnamefont {Egor}\ \bibnamefont
  {Nikulin}}, \bibinfo {author} {\bibfnamefont {Dmitry}\ \bibnamefont
  {Mylnikov}}, \bibinfo {author} {\bibfnamefont {Denis}\ \bibnamefont
  {Bandurin}}, \ and\ \bibinfo {author} {\bibfnamefont {Dmitry}\ \bibnamefont
  {Svintsov}},\ }\bibfield  {title} {\enquote {\bibinfo {title} {{Edge
  diffraction, plasmon launching, and universal absorption enhancement in
  two-dimensional junctions}},}\ }\href {\doibase 10.1103/PhysRevB.103.085306}
  {\bibfield  {journal} {\bibinfo  {journal} {Physical Review B}\ }\textbf
  {\bibinfo {volume} {103}},\ \bibinfo {pages} {085306} (\bibinfo {year}
  {2021})}\BibitemShut {NoStop}%
\bibitem [{\citenamefont {Artmann}(1948)}]{Artman_GH_theory}%
  \BibitemOpen
  \bibfield  {author} {\bibinfo {author} {\bibfnamefont {Kurt}\ \bibnamefont
  {Artmann}},\ }\bibfield  {title} {\enquote {\bibinfo {title} {Berechnung der
  seitenversetzung des totalreflektierten strahles},}\ }\href {\doibase
  https://doi.org/10.1002/andp.19484370108} {\bibfield  {journal} {\bibinfo
  {journal} {Annalen der Physik}\ }\textbf {\bibinfo {volume} {437}},\ \bibinfo
  {pages} {87--102} (\bibinfo {year} {1948})}\BibitemShut {NoStop}%
\bibitem [{\citenamefont {Mikhailov}\ and\ \citenamefont
  {Volkov}(1992)}]{IE_Plasmons}%
  \BibitemOpen
  \bibfield  {author} {\bibinfo {author} {\bibfnamefont {S~A}\ \bibnamefont
  {Mikhailov}}\ and\ \bibinfo {author} {\bibfnamefont {V~A}\ \bibnamefont
  {Volkov}},\ }\bibfield  {title} {\enquote {\bibinfo {title} {Inter-edge
  magnetoplasmons in inhomogeneous two-dimensional electron systems},}\ }\href
  {\doibase 10.1088/0953-8984/4/31/005} {\bibfield  {journal} {\bibinfo
  {journal} {Journal of Physics: Condensed Matter}\ }\textbf {\bibinfo {volume}
  {4}},\ \bibinfo {pages} {6523} (\bibinfo {year} {1992})}\BibitemShut
  {NoStop}%
\bibitem [{\citenamefont {Rodrigo}\ \emph {et~al.}(2015)\citenamefont
  {Rodrigo}, \citenamefont {Limaj}, \citenamefont {Janner}, \citenamefont
  {Etezadi}, \citenamefont {de~Abajo}, \citenamefont {Pruneri},\ and\
  \citenamefont {Altug}}]{Graphene-biosensor}%
  \BibitemOpen
  \bibfield  {author} {\bibinfo {author} {\bibfnamefont {Daniel}\ \bibnamefont
  {Rodrigo}}, \bibinfo {author} {\bibfnamefont {Odeta}\ \bibnamefont {Limaj}},
  \bibinfo {author} {\bibfnamefont {Davide}\ \bibnamefont {Janner}}, \bibinfo
  {author} {\bibfnamefont {Dordaneh}\ \bibnamefont {Etezadi}}, \bibinfo
  {author} {\bibfnamefont {F.~Javier~García}\ \bibnamefont {de~Abajo}},
  \bibinfo {author} {\bibfnamefont {Valerio}\ \bibnamefont {Pruneri}}, \ and\
  \bibinfo {author} {\bibfnamefont {Hatice}\ \bibnamefont {Altug}},\ }\bibfield
   {title} {\enquote {\bibinfo {title} {Mid-infrared plasmonic biosensing with
  graphene},}\ }\href {\doibase 10.1126/science.aab2051} {\bibfield  {journal}
  {\bibinfo  {journal} {Science}\ }\textbf {\bibinfo {volume} {349}},\ \bibinfo
  {pages} {165--168} (\bibinfo {year} {2015})}\BibitemShut {NoStop}%
\bibitem [{\citenamefont {Dyakonov}\ and\ \citenamefont
  {Shur}(1993)}]{DS_Instab}%
  \BibitemOpen
  \bibfield  {author} {\bibinfo {author} {\bibfnamefont {Michael}\ \bibnamefont
  {Dyakonov}}\ and\ \bibinfo {author} {\bibfnamefont {Michael}\ \bibnamefont
  {Shur}},\ }\bibfield  {title} {\enquote {\bibinfo {title} {Shallow water
  analogy for a ballistic field effect transistor: New mechanism of plasma wave
  generation by dc current},}\ }\href {\doibase 10.1103/PhysRevLett.71.2465}
  {\bibfield  {journal} {\bibinfo  {journal} {Phys. Rev. Lett.}\ }\textbf
  {\bibinfo {volume} {71}},\ \bibinfo {pages} {2465--2468} (\bibinfo {year}
  {1993})}\BibitemShut {NoStop}%
\bibitem [{\citenamefont {Zonetti}\ \emph {et~al.}(2021)\citenamefont
  {Zonetti}, \citenamefont {Siaber}, \citenamefont {Cunningham},\ and\
  \citenamefont {Sydoruk}}]{Sydoruk_amplifying}%
  \BibitemOpen
  \bibfield  {author} {\bibinfo {author} {\bibfnamefont {S.}~\bibnamefont
  {Zonetti}}, \bibinfo {author} {\bibfnamefont {S.}~\bibnamefont {Siaber}},
  \bibinfo {author} {\bibfnamefont {J.~E.}\ \bibnamefont {Cunningham}}, \ and\
  \bibinfo {author} {\bibfnamefont {O.}~\bibnamefont {Sydoruk}},\ }\bibfield
  {title} {\enquote {\bibinfo {title} {{Scattering-induced amplification of
  two-dimensional plasmons: Electromagnetic modeling}},}\ }\href {\doibase
  10.1063/5.0050053} {\bibfield  {journal} {\bibinfo  {journal} {Journal of
  Applied Physics}\ }\textbf {\bibinfo {volume} {129}} (\bibinfo {year}
  {2021}),\ 10.1063/5.0050053}\BibitemShut {NoStop}%
\bibitem [{\citenamefont {Cohen}\ and\ \citenamefont
  {Goldstein}(2018)}]{Cohen}%
  \BibitemOpen
  \bibfield  {author} {\bibinfo {author} {\bibfnamefont {Roie}\ \bibnamefont
  {Cohen}}\ and\ \bibinfo {author} {\bibfnamefont {Moshe}\ \bibnamefont
  {Goldstein}},\ }\bibfield  {title} {\enquote {\bibinfo {title} {Hall and
  dissipative viscosity effects on edge magnetoplasmons},}\ }\href {\doibase
  10.1103/PhysRevB.98.235103} {\bibfield  {journal} {\bibinfo  {journal} {Phys.
  Rev. B}\ }\textbf {\bibinfo {volume} {98}},\ \bibinfo {pages} {235103}
  (\bibinfo {year} {2018})}\BibitemShut {NoStop}%
\bibitem [{\citenamefont {Sydoruk}\ \emph {et~al.}(2015)\citenamefont
  {Sydoruk}, \citenamefont {Choonee},\ and\ \citenamefont
  {Dyer}}]{Sydoruk_gated-nongated}%
  \BibitemOpen
  \bibfield  {author} {\bibinfo {author} {\bibfnamefont {Oleksiy}\ \bibnamefont
  {Sydoruk}}, \bibinfo {author} {\bibfnamefont {Kaushal}\ \bibnamefont
  {Choonee}}, \ and\ \bibinfo {author} {\bibfnamefont {Gregory~C.}\
  \bibnamefont {Dyer}},\ }\bibfield  {title} {\enquote {\bibinfo {title}
  {Transmission and reflection of terahertz plasmons in two-dimensional
  plasmonic devices},}\ }\href {\doibase 10.1109/TTHZ.2015.2405919} {\bibfield
  {journal} {\bibinfo  {journal} {IEEE Transactions on Terahertz Science and
  Technology}\ }\textbf {\bibinfo {volume} {5}},\ \bibinfo {pages} {486--496}
  (\bibinfo {year} {2015})}\BibitemShut {NoStop}%
\bibitem [{\citenamefont {Petrov}\ \emph {et~al.}(2016)\citenamefont {Petrov},
  \citenamefont {Svintsov}, \citenamefont {Rudenko}, \citenamefont {Ryzhii},\
  and\ \citenamefont {Shur}}]{Partly-gated}%
  \BibitemOpen
  \bibfield  {author} {\bibinfo {author} {\bibfnamefont {Aleksandr~S.}\
  \bibnamefont {Petrov}}, \bibinfo {author} {\bibfnamefont {D.}~\bibnamefont
  {Svintsov}}, \bibinfo {author} {\bibfnamefont {M.}~\bibnamefont {Rudenko}},
  \bibinfo {author} {\bibfnamefont {V.}~\bibnamefont {Ryzhii}}, \ and\ \bibinfo
  {author} {\bibfnamefont {M.~S.}\ \bibnamefont {Shur}},\ }\bibfield  {title}
  {\enquote {\bibinfo {title} {Plasma instability of 2d electrons in a field
  effect transistor with a partly gated channel},}\ }\href {\doibase
  10.1142/S0129156416400152} {\bibfield  {journal} {\bibinfo  {journal}
  {International Journal of High Speed Electronics and Systems}\ }\textbf
  {\bibinfo {volume} {25}},\ \bibinfo {pages} {1640015} (\bibinfo {year}
  {2016})}\BibitemShut {NoStop}%
\bibitem [{\citenamefont {Aizin}\ and\ \citenamefont {Dyer}(2012)}]{Aizin2012}%
  \BibitemOpen
  \bibfield  {author} {\bibinfo {author} {\bibfnamefont {Gregory~R.}\
  \bibnamefont {Aizin}}\ and\ \bibinfo {author} {\bibfnamefont {Gregory~C.}\
  \bibnamefont {Dyer}},\ }\bibfield  {title} {\enquote {\bibinfo {title}
  {{Transmission line theory of collective plasma excitations in periodic
  two-dimensional electron systems: Finite plasmonic crystals and Tamm
  states}},}\ }\href {\doibase 10.1103/PhysRevB.86.235316} {\bibfield
  {journal} {\bibinfo  {journal} {Physical Review B}\ }\textbf {\bibinfo
  {volume} {86}},\ \bibinfo {pages} {235316} (\bibinfo {year}
  {2012})}\BibitemShut {NoStop}%
\bibitem [{\citenamefont {Shuvaev}\ \emph {et~al.}(2022)\citenamefont
  {Shuvaev}, \citenamefont {Dzhikirba}, \citenamefont {Astrakhantseva},
  \citenamefont {Gusikhin}, \citenamefont {Kukushkin},\ and\ \citenamefont
  {Muravev}}]{Muravev-grating}%
  \BibitemOpen
  \bibfield  {author} {\bibinfo {author} {\bibfnamefont {A.}~\bibnamefont
  {Shuvaev}}, \bibinfo {author} {\bibfnamefont {K.~R.}\ \bibnamefont
  {Dzhikirba}}, \bibinfo {author} {\bibfnamefont {A.~S.}\ \bibnamefont
  {Astrakhantseva}}, \bibinfo {author} {\bibfnamefont {P.~A.}\ \bibnamefont
  {Gusikhin}}, \bibinfo {author} {\bibfnamefont {I.~V.}\ \bibnamefont
  {Kukushkin}}, \ and\ \bibinfo {author} {\bibfnamefont {V.~M.}\ \bibnamefont
  {Muravev}},\ }\bibfield  {title} {\enquote {\bibinfo {title} {Plasmonic
  metasurface created by a grating of two-dimensional electron strips on a
  substrate},}\ }\href {\doibase 10.1103/PhysRevB.106.L161411} {\bibfield
  {journal} {\bibinfo  {journal} {Phys. Rev. B}\ }\textbf {\bibinfo {volume}
  {106}},\ \bibinfo {pages} {L161411} (\bibinfo {year} {2022})}\BibitemShut
  {NoStop}%
\bibitem [{\citenamefont {Dyer}\ \emph {et~al.}(2013)\citenamefont {Dyer},
  \citenamefont {Aizin}, \citenamefont {Allen}, \citenamefont {Grine},
  \citenamefont {Bethke}, \citenamefont {Reno},\ and\ \citenamefont
  {Shaner}}]{Dyer2013}%
  \BibitemOpen
  \bibfield  {author} {\bibinfo {author} {\bibfnamefont {Gregory~C.}\
  \bibnamefont {Dyer}}, \bibinfo {author} {\bibfnamefont {Gregory~R.}\
  \bibnamefont {Aizin}}, \bibinfo {author} {\bibfnamefont {S.~James}\
  \bibnamefont {Allen}}, \bibinfo {author} {\bibfnamefont {Albert~D.}\
  \bibnamefont {Grine}}, \bibinfo {author} {\bibfnamefont {Don}\ \bibnamefont
  {Bethke}}, \bibinfo {author} {\bibfnamefont {John~L.}\ \bibnamefont {Reno}},
  \ and\ \bibinfo {author} {\bibfnamefont {Eric~A.}\ \bibnamefont {Shaner}},\
  }\bibfield  {title} {\enquote {\bibinfo {title} {{Induced transparency by
  coupling of Tamm and defect states in tunable terahertz plasmonic
  crystals}},}\ }\href {\doibase 10.1038/nphoton.2013.252} {\bibfield
  {journal} {\bibinfo  {journal} {Nature Photonics}\ }\textbf {\bibinfo
  {volume} {7}},\ \bibinfo {pages} {925--930} (\bibinfo {year}
  {2013})}\BibitemShut {NoStop}%
\end{thebibliography}%
\appendix
\begin{widetext}
\section{Wiener-Hopf method}
The scattering equation, after taking the Fourier-transform with respect to y-direction, reads as
\begin{equation}
  {{\varphi }_{{{q}_{y}}}}\left( x \right)={{\varphi }_{\text{source}}}\left( x \right)+\frac{1}{i\omega }\int{d\mathrm{{r}'}{{G}_{{{q}_{y}}}}\left( x-{x}' \right)\left[ -q_{y}^{2}\sigma \left( {{x}'} \right)+\frac{\partial }{\partial {x}'}\sigma \left( {{x}'} \right)\frac{\partial }{\partial {x}'} \right]{{\varphi }_{{{q}_{y}}}}\left( {{x}'} \right)}  
\end{equation}
where the source term has now re-emerged due to the presence of the incident wave
\begin{equation}
{{\varphi }_{\text{source}}}\left( x \right)={{\varphi }_{i}}\left[ \frac{{{\sigma }_{L}}}{i\omega }\int\limits_{-\infty }^{0}{d{x}'{{G}_{{{q}_{y}}}}\left( x-{x}' \right)\left[ -q_{y}^{2}+\frac{{{\partial }^{2}}}{\partial {{{{x}'}}^{2}}} \right]{{e}^{i{{q}_{i}}{x}'}}}-{{e}^{i{{q}_{i}}x}} \right].
\end{equation}
It is possible to observe that ${{\varphi }_{\text{source}}}\left( x \right)$ is confined in space, i.e. decays at $\left| x \right|\to \infty $. The confinement persists even if we recall that ${{q}_{i}}$ has a positive imaginary part. We now take the second Fourier transform with respect to $x$-axis:
\begin{equation}
\label{Eq-FTransformed}
 {{\varepsilon }_{L}}\left( {{q}_{x}} \right)\left[ {{\varphi }_{i}}\left( {{q}_{x}} \right)+{{\varphi }_{L}}\left( {{q}_{x}} \right) \right]+{{\varepsilon }_{R}}\left( {{q}_{x}} \right){{\varphi }_{R}}\left( {{q}_{x}} \right)=\frac{{{q}_{x}}}{\omega }G\left( q \right)\left[ {{\sigma }_{R}}-{{\sigma }_{L}} \right]{{\varphi }_{{{q}_{y}}}}\left( 0 \right),
\end{equation}
where we have split the potential into ‘left’ and ‘right’ parts ${{\varphi }_{{{q}_{y}}}}\left( x \right)={{\varphi }_{L}}\left( x \right)\theta \left( -x \right)+{{\varphi }_{R}}\left( x \right)\theta \left( x \right)$. The transform of the incident wave is
\begin{equation}
\label{Phii-transform}
{{\varphi }_{i}}\left( {{q}_{x}} \right)=\frac{i{{\varphi }_{i}}}{{{q}_{x}}-{{q}_{i}}}.
\end{equation}
Importantly, the pole of (\ref{Phii-transform}) is compensated by zero of ${{\varepsilon }_{L}}\left( {{q}_{x}} \right)$ in (\ref{Eq-FTransformed}), which can be linked to the finiteness of ${{\varphi }_{\text{source}}}\left( x \right)$ at infinity. 

Before proceeding to actual splitting, we recast the right-hand side of (\ref{Eq-FTransformed}) in terms of dielectric constants using the definition 
\begin{equation}
{{\varepsilon }_{\alpha }}\left( {{q}_{x}} \right)=1+\frac{i{{\sigma }_{\alpha }}}{\omega }{{q}^{2}}G\left( q \right).
\end{equation}
Using this, we have identically:
\begin{equation}
\frac{{{q}_{x}}}{\omega }G\left( q \right)\left[ {{\sigma }_{R}}-{{\sigma }_{L}} \right]{{\varphi }_{{{q}_{y}}}}\left( 0 \right)=\frac{i{{q}_{x}}}{q_{x}^{2}+q_{y}^{2}}\left[ {{\varepsilon }_{L}}-{{\varepsilon }_{R}} \right]{{\varphi }_{{{q}_{y}}}}\left( 0 \right)
\end{equation}
We can proceed to actual splitting now. We start from the decomposition of dielectric function 
\begin{equation}
{{\varepsilon }_{\alpha }}\left( {{q}_{x}} \right)=\varepsilon _{\alpha }^{+}\left( {{q}_{x}} \right)\varepsilon _{\alpha }^{-}\left( {{q}_{x}} \right)
\end{equation}
into the functions analytic and free of zeros in the upper (+) and lower (-) half-planes, which will be described in the next section. At this stage, we just assume that such representation is known. From our definition of Fourier transform it follows that left and right transformed potentials are analytic in the upper and lower half-planes, respectively. We now re-arrange the terms to group ${{\varphi }_{L}}\left( {{q}_{x}} \right)$ with $\varepsilon _{\alpha }^{+}\left( {{q}_{x}} \right)$, and ${{\varphi }_{R}}\left( {{q}_{x}} \right)$ with $\varepsilon _{\alpha }^{-}\left( {{q}_{x}} \right)$. This leads to
\begin{multline}
    {{M}_{+}}\left( {{q}_{x}} \right)\left[ {{\varphi }_{i}}\left( {{q}_{x}} \right)+{{\varphi }_{L}}\left( {{q}_{x}} \right) \right]+{{M}_{-}}\left( {{q}_{x}} \right){{\varphi }_{R}}\left( {{q}_{x}} \right)=\\
    -i{{\varphi }_{{{q}_{y}}}}\left( 0 \right)\frac{{{q}_{x}}}{q_{x}^{2}+q_{y}^{2}}\left[ {{M}_{-}}\left( {{q}_{x}} \right)+{{M}_{+}}\left( {{q}_{x}} \right) \right],
\end{multline}
Splitting of the right-hand side into the functions analytic in UHP and LHP is achieved by noting that
\begin{equation}
\frac{{{q}_{x}}}{q_{x}^{2}+q_{y}^{2}}=\frac{1}{2}\left( \frac{1}{{{q}_{x}}+i{{q}_{y}}}+\frac{1}{{{q}_{x}}-i{{q}_{y}}} \right).
\end{equation}

Further on, it is helpful to use the ‘pole removal’ trick to cure the functions which are analytic neither in UHP nor in the LHP. For example, in the left-hand side ${{M}_{+}}\left( {{q}_{x}} \right){{\varphi }_{i}}\left( {{q}_{x}} \right)$ has a simple pole in the UHP coming from the incident wave, while ${{M}_{+}}\left( {{q}_{x}} \right)$ may have poles in the LHP. We split this term as
\begin{equation}
   {{M}_{+}}\left( {{q}_{x}} \right){{\varphi }_{i}}\left( {{q}_{x}} \right)\equiv {{M}_{+}}\left( {{q}_{x}} \right)\frac{i{{\varphi }_{i}}}{{{q}_{x}}-{{q}_{i}}}=\underbrace{\left[ {{M}_{+}}\left( {{q}_{x}} \right)-{{M}_{+}}\left( {{q}_{i}} \right) \right]\frac{i{{\varphi }_{i}}}{{{q}_{x}}-{{q}_{i}}}}_{UHP\ analytic}+\underbrace{{{M}_{+}}\left( {{q}_{i}} \right)\frac{i{{\varphi }_{i}}}{{{q}_{x}}-{{q}_{i}}}}_{LHP\ analytic}. 
\end{equation}

Similar trick should be done several times upon splitting of the right-hand side. Proceeding this way, we get a split equation
\begin{multline}
 \left[ {{M}_{+}}\left( {{q}_{x}} \right)-{{M}_{+}}\left( {{q}_{i}} \right) \right]{{\varphi }_{i}}\left( {{q}_{x}} \right)+{{M}_{+}}\left( {{q}_{x}} \right){{\varphi }_{L}}\left( {{q}_{x}} \right)-i\frac{{{\varphi }_{{{q}_{y}}}}\left( 0 \right)}{2}{{L}_{+}}\left( {{q}_{x}} \right)= \\ 
 -{{M}_{+}}\left( {{q}_{i}} \right){{\varphi }_{i}}\left( {{q}_{x}} \right)-{{M}_{-}}\left( {{q}_{x}} \right){{\varphi }_{R}}\left( {{q}_{x}} \right)+i\frac{{{\varphi }_{{{q}_{y}}}}\left( 0 \right)}{2}{{L}_{-}}\left( {{q}_{x}} \right)  
\end{multline}
where
\begin{equation}
    {{L}_{\pm }}\left( {{q}_{x}} \right)=\pm \frac{{{M}_{\pm }}\left( {{q}_{x}} \right)}{{{q}_{x}}\pm i{{q}_{_{y}}}}\pm \frac{{{M}_{\pm }}\left( {{q}_{x}} \right)-{{M}_{\pm }}\left( \pm i{{q}_{y}} \right)}{{{q}_{x}}\mp i{{q}_{_{y}}}}\mp \frac{{{M}_{\mp }}\left( \mp i{{q}_{y}} \right)}{{{q}_{x}}\pm i{{q}_{_{y}}}}.
\end{equation}

\section{Splitting of dielectric function}
Splitting of dielectric function can be generally achieved with
\begin{equation}
\label{Eq-direct-splitting}
\varepsilon _{\alpha }^{\pm }({{q}_{x}})=\exp \left\{ \pm \frac{1}{2\pi i}\int_{-\infty }^{+\infty }{\frac{\ln {{\varepsilon }_{\alpha }}(u)du}{u-{{q}_{x}}\pm i{{0}^{+}}}} \right\}.
\end{equation}
The branch cuts of the logarithm of complex-valued function ${{\varepsilon }_{\alpha }}(u)$ should be chosen not to cross the real axis. A direct numerical evaluation of integral (\ref{Eq-direct-splitting}) is time-consuming due to the presence of zeros of ${{\varepsilon }_{\alpha }}(u)$ in the immediate vicinity of real axis. They occur at $\pm \sqrt{q_{p\alpha }^{2}-q_{y}^{2}}$ and may have very small imaginary part, which makes $\ln {{\varepsilon }_{\alpha }}(u)$ very large in the vicinity of these points. To avoid this complexity, we rewrite the dielectric function as:
\begin{equation}
\label{Eq-approximants}
{{\varepsilon }_{\alpha }}({{q}_{x}})={{\varepsilon }_{\text{app}}}({{q}_{x}})\frac{{{\varepsilon }_{\alpha }}({{q}_{x}})}{{{\varepsilon }_{\text{app}}}({{q}_{x}})},
\end{equation}
where the approximant of the dielectric function ${{\varepsilon }_{\text{app}}}({{q}_{x}})$ should have the same zeros as ${{\varepsilon }_{\alpha }}({{q}_{x}})$, and be a polynomial function of wave vector. After that, splitting of ${{\varepsilon }_{\text{app}}}({{q}_{x}})$ would be achieved automatically, while the remainder can be factorized with (\ref{Eq-direct-splitting}). The emerging principal-value integral will be well-behaved, as ${{\varepsilon }_{\alpha }}({{q}_{x}})/{{\varepsilon }_{\text{app}}}({{q}_{x}})$ is zero-free.

We further limit ourselves to the splitting of dielectric function for non-gated 2DES
\begin{equation}
    {{\varepsilon }_{\alpha }}({{q}_{x}})=1+\frac{2\pi i{{\sigma }_{\alpha }}}{\omega }\sqrt{q_{x}^{2}+q_{y}^{2}}=1-\sqrt{\frac{q_{x}^{2}+q_{y}^{2}}{q_{p\alpha }^{2}}}
\end{equation}
where we have introduced the plasma wave vector
\begin{equation}
    q_{p\alpha }^{2}=\frac{i\omega }{2\pi {{\sigma }_{\alpha }}}.
\end{equation}
The approximant for such dielectric function is
\begin{equation}
    {{\varepsilon }_{\text{app}}}({{q}_{x}})=1-\frac{q_{x}^{2}+q_{y}^{2}}{q_{p\alpha }^{2}}=\underbrace{\frac{\sqrt{q_{p\alpha }^{2}-q_{y}^{2}}-{{q}_{x}}}{{{q}_{p\alpha }}}}_{\varepsilon _{app}^{-}\left( {{q}_{x}} \right),\,LHP\ analytic}\underbrace{\frac{{{q}_{x}}+\sqrt{q_{p\alpha }^{2}-q_{y}^{2}}}{{{q}_{p\alpha }}}}_{\varepsilon _{app}^{+}\left( {{q}_{x}} \right),\,UHP\ analytic}.
\end{equation}

Now, to factorize (\ref{Eq-approximants}) we can use
\begin{equation}
    \varepsilon _{\alpha }^{\pm }({{q}_{x}})=\varepsilon _{app}^{\pm }({{q}_{x}})\exp \left\{ \pm \frac{1}{2\pi i}\int_{-\infty }^{+\infty }{\frac{du\,\ln \frac{{{\varepsilon }_{\alpha }}(u)}{{{\varepsilon }_{\text{app}}}(u)}}{u-{{q}_{x}}\pm i{{0}^{+}}}} \right\}=\frac{\pm {{q}_{x}}+\sqrt{q_{p\alpha }^{2}-q_{y}^{2}}}{{{q}_{p\alpha }}}\exp \left\{ \pm \frac{1}{2\pi i}\int_{-\infty }^{+\infty }{\frac{du\,\ln \left[ 1+\sqrt{\frac{{{u}^{2}}+q_{y}^{2}}{q_{p\alpha }^{2}}} \right]}{u-{{q}_{x}}\pm i{{0}^{+}}}} \right\},
\end{equation}

To complete the discussion, we write down some approximate formulas useful for evaluations. We note that dielectric function is determined only by two dimensionless parameters $a={{q}_{x}}/{{q}_{y}}$ and $b={{q}_{p\alpha }}/{{q}_{y}}>1$:
\begin{equation}
    \varepsilon _{\alpha }^{\pm }=\frac{\pm a+\sqrt{{{b}^{2}}-1}}{b}\exp \left\{ \pm \frac{1}{2\pi i}I\left( a,b \right) \right\},I\left( a,b \right)=\int_{-\infty }^{+\infty }{\frac{du\,\ln \left[ 1+\frac{1}{b}\sqrt{{{u}^{2}}+1} \right]}{u-a}}.
\end{equation}

Upon evaluation of $\varepsilon _{\alpha }^{+}$ we should treat $a={a}'+i{{a}'}'$, ${{a}'}'>0$, while the $u$-integration is now along the real axis. The derivative of integral $I\left( a,b \right)$ can be evaluated via elementary functions:
\begin{equation}
\frac{\partial I\left( a,b \right)}{\partial a}=\int\limits_{-\infty }^{+\infty }{\frac{udu}{\sqrt{{{u}^{2}}+1}(u-a)\left( b+\sqrt{{{u}^{2}}+1} \right)}}=\frac{-2\sqrt{{{b}^{2}}-1}{\rm arccosh} b+\frac{a\left( 2b {\rm arcsinh} a+i\pi \left( \sqrt{{{a}^{2}}+1}-b \right) \right)}{\sqrt{{{a}^{2}}+1}}}{{{a}^{2}}-{{b}^{2}}+1} 
\end{equation}

The branch cuts of inverse hyperbolic functions ${\rm arccosh} z$ and ${\rm arcsinh} z$ should run from $\pm i$ to $\pm i\infty $. Restoration of $I\left( a,b \right)$ from its $a$-derivative is possible analytically, but involves dilogarithm functions. Here, we present only a specific value required for reflection and transmission computations, namely $\varepsilon _{\alpha }^{+}\left( i{{q}_{y}} \right)$. After simple transformations, we find
\begin{equation}
    \varepsilon _{\alpha }^{+}\left( {{q}_{x}}=i{{q}_{y}} \right)=\frac{i+\sqrt{{{b}^{2}}-1}}{b}\exp \left\{ \frac{1}{2\pi }G\left( b \right) \right\},G\left( b \right)=\int_{-\infty }^{+\infty }{\frac{du\,\ln \left[ 1+\frac{1}{b}\sqrt{{{u}^{2}}+1} \right]}{{{u}^{2}}+1}}=\int_{-\infty }^{+\infty }{\frac{dx\,\ln \left[ 1+\frac{\cosh x}{b} \right]}{\cosh x}}.
\end{equation}
The derivative of the integral is simply evaluated
\begin{equation}
\frac{\partial G}{\partial b}=-\frac{2\ln \left( \sqrt{{{b}^{2}}-1}+b \right)}{b\sqrt{{{b}^{2}}-1}}
\end{equation}

Restoration of the full integral is here relatively simple
\begin{equation}
    G\left( b \right)=2i\left\{ \text{L}{{\text{i}}_{2}}\left( i\left( \sqrt{{{b}^{-2}}-1}-{{b}^{-1}} \right) \right)-\text{L}{{\text{i}}_{2}}\left( \frac{i}{{{b}^{-1}}+\sqrt{{{b}^{-2}}-1}} \right) \right\}+4\ln \left( \sqrt{{{b}^{-2}}-1}+{{b}^{-1}} \right){\rm arccot} \left( \sqrt{{{b}^{-2}}-1}+{{b}^{-1}} \right).
\end{equation}

\end{widetext}

\end{document}